\documentclass[AMA,STIX1COL]{WileyNJD-v2}
\usepackage{moreverb}
\usepackage{xcolor}
\usepackage{bbm}
\usepackage{amsmath}
\usepackage{setspace}
\usepackage{subcaption} 
\usepackage{lscape}

\hypersetup{hidelinks}
\newcommand\BibTeX{{\rmfamily B\kern-.05em \textsc{i\kern-.025em b}\kern-.08em
T\kern-.1667em\lower.7ex\hbox{E}\kern-.125emX}}

\articletype{Research Article}%

\received{<day> <Month>, <year>}
\revised{<day> <Month>, <year>}
\accepted{<day> <Month>, <year>}

\def\negr#1{\mbox{\boldmath$#1$}}
\newcommand{\vect}[1]{\mathbf{#1}}
\def\E{{\mathbb E}} 
\newcommand{\addeq}{ \overset{\textrm{\tiny{+}}}{\approx}}

\begin{document}
\doublespacing
\title{Fast variational Bayesian inference for correlated survival data: an application to invasive mechanical ventilation duration analysis} 

\author[1]{Chengqian Xian}

\author[1]{Camila P.E. de Souza}

\author[1]{Wenqing He}

\author[1,2]{Felipe F. Rodrigues}

\author[2]{Renfang Tian}

\authormark{Chengqian Xian \textsc{et al}}

\address[1]{\orgdiv{Department of Statistical and Actuarial Sciences}, \orgname{Western University}, \orgaddress{\state{London, Ontario}, \country{Canada}}}

\address[2]{\orgdiv{School of Management, Economics, and Mathematics}, \orgname{King's University College at Western University}, \orgaddress{\state{London, Ontario}, \country{Canada}}}

\corres{Chengqian Xian, Department of Statistical and Actuarial Sciences, Western University, 1151 Richmond Street, London, Ontario, Canada, N6A 5B7. \email{cxian3@uwo.ca}}


\abstract[Abstract]{Correlated survival data are prevalent in various clinical settings and have been extensively discussed in literature. One of the most common types of correlated survival data is clustered survival data, where the survival times from individuals in a cluster are associated. Our study is motivated by invasive mechanical ventilation data from different intensive care units (ICUs) in Ontario, Canada, forming multiple clusters. The survival times from patients within the same ICU cluster are correlated. To address this association, we introduce a shared frailty log-logistic accelerated failure time model that accounts for intra-cluster correlation through a cluster-specific random intercept. We present a novel, fast variational Bayes (VB) algorithm for parameter inference and evaluate its performance using simulation studies varying the number of clusters and their sizes. We further compare the performance of our proposed VB algorithm with the h-likelihood method and a Markov Chain Monte Carlo (MCMC) algorithm. The proposed algorithm delivers satisfactory results and demonstrates computational efficiency over the MCMC algorithm. We apply our method to the ICU ventilation data from Ontario to investigate the ICU site random effect on ventilation duration.}



\maketitle

\section{Introduction}\label{sec1}
Correlated survival data are commonplace in various research contexts and have been extensively studied in the literature \citep{Luo_2013, Honerkamp_2016, Liu_207}. One such context of correlated survival data is observed in clustered survival data, which are derived from multiple entities such as families or hospitals. Within each cluster, survival data exhibit correlation because of the shared environmental factors, and random effects are often introduced to capture the shared characteristics within the cluster. Conditional on the random effects (frailty), the survival times within a cluster can be assumed independent, which leads to a shared frailty model to account for cluster-level uncertainty \citep{Hougaard_1995, Hanagal_2011, Gorfine_2023}.

Our research is motivated by data from intensive care units (ICUs) across multiple clinical centers in Ontario, Canada. The ICU data were provided by the Critical Care Information System (CCIS) Ontario database. In Canadian ICUs, invasive mechanical ventilation is prevalent, with approximately one third of patients requiring ventilation during their ICU stay \cite{ICU_2020}. The duration of ventilation has significant implications for clinical outcomes and is associated with an increased risk of complications \cite{ICU_2020}. Analysis on the ventilation duration for ICU patients utilizing the associated risk factors, such as patient categories (e.g., medical or surgical), admission diagnoses, and patient severity, can help determine the number of beds with ventilators and therefore, support capacity planning and effective clinical resource management. The severity of patients in ICU is often assessed using the Multiple Organ Dysfunction Score (MODS) \cite{Marshall_1995} that evaluates organ function. Kobara et al. \cite{Kobara_2023} conducted a study on ventilation duration for ICU patients from Ontario using a survival analysis framework. In their study, patient ventilation duration is considered as a time-to-event (survival) outcome and parametric accelerated failure time (AFT) models are applied to predict the ventilation time. If a patient is transferred to another facility without subsequent follow-up information, the ventilation time is considered right censored. Kobara et al. found that the log-logistic AFT model well describes the association between risk factors and patients’ ventilation duration in Ontario ICUs. However, Kobara et al. did not consider a possible correlation of ventilation duration times among patients within the same ICU. Previous studies have indicated that patient outcomes within the same ICU site may be correlated, and ignoring this hierarchical structure can result in flawed prediction models \cite{Burgess_2000, Glance_2003}. 

As aforementioned, shared frailty can be used to accommodate cluster-level association among patients within the cluster. Gorfine and Zucker \citep{Gorfine_2023} recently provided a comprehensive review of shared frailty methods for complex survival data. To incorporate risk factors in survival data, the Cox proportional hazards (PH) model with a multiplicative shared frailty on the hazard rates has been widely developed \citep{Gorfine_2023}. As an alternative to the Cox PH model, AFT models offer an intuitive interpretation of covariate effects on survival time. Lambert et al. \citep{Lambert_2004} developed shared frailty AFT models with different parametric distributions assumed for the survival time and conducted maximum likelihood estimation by integrating out the unobserved frailties. They empirically found that the choice of distribution for the shared frailty is not critical, recommending the normal distribution. Do Ha et al. \citep{Do_2002, Do_2017} considered a shared frailty log-normal AFT model and used the hierarchical-likelihood (h-likelihood) approach to estimate the model parameters. The h-likelihood method obtains fixed effects estimates by maximizing the h-likelihood, while utilizing the restricted maximum likelihood estimate for the estimation of the variance of the shared frailty \cite{Do_2017}. An R package, \texttt{frailtyHL} \citep{Do_2012} has been developed to implement the h-likelihood estimation in the shared frailty log-normal AFT model. Building on this, Park and Do Ha \citep{Park_2019} introduced a penalized variable selection technique for the shared frailty log-normal AFT model. Additionally, Zhou et al. \citep{Zhou_2017} developed a Markov chain Monte Carlo (MCMC)-based algorithm, called \textit{survregBayes}, to estimate the shared frailty AFT model, and an R package, \texttt{spBayesSurv} \citep{Zhou_2020}, is available for its implementation. To our knowledge, there is no work on variational Bayesian inference for shared frailty AFT models. 

As an alternative to MCMC methods, which are the gold standard for obtaining posterior distributions under a Bayesian framework, variational inference (VI) has gained popularity due to its favourable results and lower computational cost than MCMC. Recently, several types of VI algorithms have been developed, including mean-field VI \citep{Bishop_2006}, stochastic VI \citep{Hoffman_2013}, and black-box VI \citep{Ranganath_2014}. A special case of mean-field VI, called mean-field variational Bayes (VB), arises when the Kullback–Leibler (KL) divergence is utilized to quantify the dissimilarity between exact and approximated posterior distributions. In addition, the approximated posterior distribution, referred to as the variational posterior, is assumed to belong to a mean-field variational family. Under the mean-field VB framework, the solutions to minimizing the KL divergence can be obtained by utilizing the coordinate ascent algorithm \cite{Bishop_2006, Jordan_1999}. Mean-field VB has been widely applied to regression models such as the generalized additive model \citep{Neville_2011}, nonparametric regression with measurement error \citep{Pham_2013}, count response semiparametric regression \citep{Luts_2015}, high-dimensional linear regression \citep{Ray_2022}, multilevel regression modelling \citep{Lee_2016}, B-spline regression mixture model for functional data clustering \citep{Xian_clu_2024}, basis selection for functional data representation \citep{daCruz_2024}, among others. Applications of other types of VI can be found in a comprehensive review of VI from a statistical perspective by Blei et al. \citep{David_2017}.

In this study, we propose a shared frailty log-logistic AFT model to account for the correlation among patient ventilation durations within the ICU sites. We develop a novel and fast mean-field VB algorithm to infer the model parameters. By applying the piece-wise approximation techniques proposed by Xian et al. \cite{Xian_2024} to avoid intractable calculations, we obtain closed-form posterior distributions. We conduct extensive simulation studies with various numbers of clusters and cluster sizes to evaluate the performance of the proposed method, and compare the performance of our VB algorithm with the h-likelihood method \citep{Do_2017} and the MCMC-based algorithm \textit{survregBayes} \citep{Zhou_2020}. Finally, we apply our methodology to investigate ventilation duration for ICU patients using the same dataset as Kobara et al. \cite{Kobara_2023}. This study was approved by the Research Ethics Review Committee at King’s University College at Western University.

The remainder of the paper is organized as follows. Section \ref{Sec:model} presents the log-logistic AFT model with a shared frailty under the Bayesian framework. We introduce our proposed VB algorithm in Section \ref{sec:algorithm}. In Section \ref{sec:simulation}, simulation studies are conducted to evaluate the performance of our VB algorithm. Section \ref{sec:real} illustrates the application of the proposed method to the ICU ventilation duration data. A discussion is provided in Section \ref{Sec:Dis}.
\section{Bayesian Log-logistic AFT model with a shared frailty}\label{Sec:model}
Let $T_{ij}$ and $C_{ij}$ be the survival and censoring times, respectively, of the $j^{th}$ subject from the $i^{th}$ group (i.e., cluster) in a sample, $i=1, ..., K$ and $j=1, ..., n_i$. Let $t_{ij}=\min(T_{ij}, C_{ij})$ and $\delta_{ij}=\mathbbm{1}(T_{ij} \leq C_{ij})$ be the subject's observed time and the indicator for right censoring, respectively. We consider a log-logistic AFT model with shared frailty (a random intercept) specified as follows:
\begin{equation}
    \log(T_{ij})=\gamma_i + \vect{X}_{ij}^T\,\negr{\beta} + b\epsilon_{ij} 
\label{AFT.model.random.intercept}
\end{equation}
where $\vect{X}_{ij}$ is a column vector with length $p, p \geq 2,$ containing $p-1$ fixed effects (covariates) and a constant one to incorporate the constant intercept (i.e., $\vect{X}_{ij} = (1, x_{ij1}, ..., x_{ij(p-1)})^T$), $\negr{\beta}$ is the corresponding vector of coefficients for the fixed effects, where $\gamma_i$ is a random intercept for the $i^{th}$ cluster and $\epsilon_{ij}$ is a random variable following a standard logistic distribution, and $b$ is a scale parameter. We further assume that $\gamma_i \overset{\mathrm{iid}}{\sim} N(0, \sigma^2_{\gamma})$ representing the discrepancy between clusters with iid denoting identically and independent distributed. The survival time $T_{ij}$ and censoring time $C_{ij}$ are assumed independent given the covariates $\vect{X}_{ij}$. Our model follows a structure similar to one presented by Robison (1991) \citep{Robinson_1991} and Nolan et al. \citep{Nolan_2020} for a Gaussian linear mixed effect model. 

In our proposed Model (\ref{AFT.model.random.intercept}), we incorporate the unknown and unobserved shared risk through cluster-specific random intercepts and estimate the model parameters, $\negr{\beta}$, $b$, and $\sigma^2_{\gamma}$ using a Bayesian framework by further assuming the following independent marginal prior distributions:

\begin{itemize}
\setlength{\itemsep}{1pt}
    \item $\negr{\beta} \sim N_p(\negr\mu_0, \sigma_0^2 \textbf{I}_{p\times p}) \; \text{with precision} \; v_0=1/\sigma_0^2$ and $\textbf{I}_{p \times p}$ being a $p \times p$ identity matrix
    \item $b \sim \text{Inverse-Gamma}\,(\alpha_0, \omega_0)$
    \item $\gamma_i \vert \sigma_{\gamma}^2 \overset{\mathrm{iid}}{\sim} N(0, \sigma^2_{\gamma}), i=1,..., K$
    \item $\sigma_{\gamma}^2\sim \text{Inverse-Gamma}\,(\lambda_0, \eta_0)$
\end{itemize}
where $\mu_0, v_0, \alpha_0$, $\omega_0$, $\lambda_0$ and $\eta_0$ are known hyperparameters (parameters of the prior distributions).

\section{Variational Bayes algorithm}\label{sec:algorithm}
In what follows, we outline our methodology for deriving a mean-field VB algorithm for Model (\ref{AFT.model.random.intercept}). We summarize the resulted VB algorithm in Algorithm \ref{VBsurvival:frailty}.

Given the observed data $\vect{D}:=\{(t_{ij}, \delta_{ij}, \vect X_{ij}), i=1, ..., K, j=1,..., n_i\}$, we denote the complete data log-likelihood by $\log p(\vect{D}, \negr{\beta}, \negr{\gamma}, b, \sigma^2_{\gamma})$, where $\negr{\gamma}=(\gamma_1, ..., \gamma_K)$. Our objective is to derive a VB algorithm to approximate the exact posterior joint distribution of $\negr{\beta}$, $b$, $\negr{\gamma}$ and $\sigma^2_{\gamma}$ given the data $\vect{D}$ by maximizing the evidence lower bound (ELBO) defined as 
\begin{eqnarray}
\text{ELBO}(q)=\E_{q}[\log p(\vect{D}, \negr{\beta}, \negr{\gamma}, b, \sigma^2_{\gamma})]-\E_{q}[\log q(\negr{\beta}, b, \negr{\gamma}, \sigma^2_{\gamma})]
\label{eq:ELBO}
\end{eqnarray}
where $q(\negr{\beta}, b, \negr{\gamma}, \sigma^2_{\gamma})$ is the approximated posterior joint distribution, which is also called the variational density, and the expectation is taken with respect to the variational density \citep{David_2017}. We consider the mean-field variational family which assumes that $q(\negr{\beta}, b, \negr{\gamma}, \sigma^2_{\gamma})=q(\negr{\beta})\, q(b)\, q(\sigma^2_{\gamma})\,\prod_{i=1}^K q({\gamma}_i)$, and apply the coordinate ascent variational inference (CAVI) algorithm \cite{Bishop_2006} to obtain each variational component (e.g., $q(\negr{\beta})$) in $q(\negr{\beta}, b, \negr{\gamma}, \sigma^2_{\gamma})$). Under our Model (\ref{AFT.model.random.intercept}), the complete data log-likelihood $\log p(\vect{D}, \negr{\beta}, \negr{\gamma}, b, \sigma^2_{\gamma})$ can be obtained by 
\begin{eqnarray}
\log p(\vect{D}, \negr{\beta}, \negr{\gamma}, b, \sigma^2_{\gamma})=\log p(\vect{D}\,\vert\,\negr{\beta}, \negr{\gamma}, b)+\log p(\negr{\beta})+\log p(b) + \sum_{i=1}^K \log p(\negr{\gamma}_i\, \vert \sigma^2_{\gamma}) + \log p(\sigma^2_{\gamma}) \label{eq:comp:dt:log:like}
\end{eqnarray}
where 
\begin{eqnarray}
{\log p(\vect{D}\,\vert\,\negr{\beta}, \negr{\gamma}, b)} = -\delta\log b +\sum_{i=1}^K\sum_{j=1}^{n_i} \Bigg[\delta_{ij} \frac{y_{ij}-\vect{X}_{ij}^T\negr{\beta}-{\gamma}_i}{b} -(1+\delta_{ij})\log\Big\{1+\exp\big(\frac{y_{ij}-\vect{X}_{ij}^T \negr{\beta}-{\gamma}_i}{b}\big)\Big\}\Bigg] \label{eq:loglike}
\end{eqnarray}
with $\delta=\sum_{i=1}^K\sum_{j=1}^{n_i}\delta_{ij}$ being the number of observed uncensored times and $y_{ij}=\log(t_{ij}), i=1, ..., K, j= 1, ..., n_i,$ being the log observed survival time.

By maximizing the ELBO, the following solutions are provided by the CAVI algorithm:
\begin{eqnarray}
\log q^*(\negr{\beta}) \addeq \E_{-\negr{\beta}}[\log p(\vect{D}\,\vert\,\negr{\beta}, \negr{\gamma}, b)+\log p(\negr{\beta})] \nonumber
\end{eqnarray}
\begin{eqnarray}
\log q^*({\gamma}_i) \addeq \E_{-{\gamma}_i}\Big[\log p(\vect{D}\,\vert\,\negr{\beta}, \negr{\gamma}, b)+\sum_{i=1}^K \log p({\gamma}_i\, \vert \sigma^2_{\gamma})\Big] \nonumber
\end{eqnarray}
\begin{eqnarray}
\log q^*(b) \addeq \E_{-b}[\log p(\vect{D}\,\vert\,\negr{\beta}, \negr{\gamma}, b)+\log p(b)] \nonumber
\end{eqnarray}
\begin{eqnarray}
\log q^*(\sigma^2_{\gamma}) \addeq \E_{-\sigma^2_{\gamma}}\Big[\sum_{i=1}^K \log p({\gamma}_i\, \vert \sigma^2_{\gamma})+\log p(\sigma^2_{\gamma})\Big]\nonumber
\end{eqnarray}
where we use $\addeq$ to denote equality up to a constant additive factor for convenience, and $-\negr{\beta}$ indicates the expectation is taken with respect to the variational density of other latent variables but $\negr{\beta}$, same for other solutions. To achieve conjugacy and tractable expectation calculation of $\log p(\vect{D}\,\vert\,\negr{\beta}, \negr{\gamma}, b)$ as specified in (\ref{eq:loglike}), we apply the proposed method by Xian et al. \citep{Xian_2024}, piecewise approximations of the function, $f(x) = \log(1+\exp(x)), x \in (-\infty, \infty)$, embedded in deriving the update equations for each parameter. As in Nolan et al. \citep{Nolan_2020} and Lee and Wand \citep{LeeCathy_2016}, we are also interested in the posterior distribution of the random effects, $q^*({\gamma}_i), i=1,...,K$, in our proposed VB framework.

\subsection{Update equation for each variational density}\label{Sec:update}
The update equations to obtain the optimal variational densities of $\negr{\beta}$, ${\gamma}_i$, $b$ and $\sigma^2_{\gamma}$ denoted by $q^*(\negr{\beta})$, $q^*({\gamma}_i)$, $q^*(b)$ and $q^*(\sigma^2_{\gamma})$, respectively, which are the corresponding approximated posterior distributions, are presented as follows within the paper while their derivations are given in the Appendix A. The calculation of expectations in the update equations are given in Section \ref{Sec:ELBO}. In the update equations, $\varphi_{ij}$, $\zeta_{ij}$, and $\rho_{ij}$ represent the piece-wise approximation coefficients proposed by Xian et al. \cite{Xian_2024} for the $j^{th}$ subject in the $i^{th}$ cluster. Detailed information regarding these coefficients is also provided in the Appendix A. 

\noindent\textbf{\textit{(1) Update equation for $q^*(\negr{\beta})$}}\vspace{0.2cm}

$q^*(\negr{\beta})$ is an $N_p(\negr{\mu}^{*}, \Sigma^*)$ where
\begin{eqnarray}
\Sigma^*= \bigg[v_0\textbf{I}_{p \times p}+2\E_{q(b)}\big(\frac{1}{b^2}\big) \sum_{i=1}^K\sum_{j=1}^{n_i}(1+\delta_{ij})\zeta_{ij}\vect{X}_{ij}\vect{X}_{ij}^T\bigg]^{-1}
\nonumber
\end{eqnarray}
and 
\begin{eqnarray}
\negr{\mu}^*=\Bigg[\bigg\{v_0\,\negr{\mu}_0^T + \sum_{i=1}^K\sum_{j=1}^{n_i}\bigg( \Big(\E_{q(b)}\big(\frac{1}{b}\big)\Big(-\delta_{ij}+(1+\delta_{ij})\rho_{ij}\Big) +2\E_{q(b)}\big(\frac{1}{b^2}\big) (1+\delta_{ij})\zeta_{ij}\big(y_{ij}-\E_{q({\gamma}_i)}({\gamma}_i)\big) \Big)\vect{X}_{ij}^T\bigg)\bigg\}\,\Sigma^{*}\Bigg]^{T}.
\nonumber
\end{eqnarray}

\noindent\textbf{\textit{(2) Update equation for $q^*({\gamma}_i)$}}\vspace{0.2cm}

$q^*({\gamma}_i)$ is an $N_l(\tau_i^*, \sigma_i^{2*})$ where
\begin{eqnarray}
\tau^*_{i}=\sigma^{2*}_{i} \sum_{j=1}^{n_i}\Big[\E_{q(b)}(\frac{1}{b})\big(-\delta_{ij}+(1+\delta_{ij})\rho_{ij}\big) +2\E_{q(b)}(\frac{1}{b^2})(1+\delta_{ij})\zeta_{ij}\big(y_{ij}-\vect{X}_{ij}^T\E_{q(\negr{\beta})}\negr{\beta}\big)\Big]  \nonumber
\end{eqnarray}
and 
\begin{eqnarray}
\sigma^{2*}_{i}= \big[\E_{q(\sigma^2_{\gamma})}\big(\frac{1}{\sigma^2_{\gamma}}\big)+2\E_{q(b)}\big(\frac{1}{b^2}\big)\sum_{j=1}^{n_i}(1+\delta_{ij})\zeta_{ij}\big]^{-1}.\nonumber
\end{eqnarray}

\noindent\textbf{\textit{(3) Update equation for $q^*(b)$}}\vspace{0.2cm}

$q^*(b)$ is an $\text{Inverse-Gamma}\,(\alpha^*, \omega^*)$ where
$\alpha^* = \alpha_0 + \delta$ and
\begin{eqnarray}
\omega^* = \omega_0-\sum_{i=1}^K\sum_{j=1}^{n_i}\Big(\delta_{ij}-(1+\delta_{ij})\varphi_{ij}\Big)\Big(y_{ij}-\vect{X}_{ij}^T\E_{q(\negr{\beta})}(\negr{\beta})-\E_{q(\gamma_i)}\gamma_i\Big).\nonumber
\end{eqnarray}

\noindent\textbf{\textit{(4) Update equation for $q^*(\sigma^2_{\gamma})$}}\vspace{0.2cm}

$q^*(\sigma^2_{\gamma})$ is an $\text{Inverse-Gamma}\,(\lambda^*, \eta^*)$ where $\lambda^*=\lambda_0 + K/2$ and
\begin{eqnarray}
\eta^* = \eta_0 + \frac{1}{2}\sum_{i=1}^K\E_{q(\gamma_i)}\gamma_i^2. \nonumber
\end{eqnarray}
\subsection{ELBO calculation}\label{Sec:ELBO}
The ELBO under Model (\ref{AFT.model.random.intercept}) is defined in (\ref{eq:ELBO}) with the complete-data log-likelihood calculated by (\ref{eq:comp:dt:log:like}) and note that $q(\negr{\beta}, b, \negr{\gamma}, \sigma^2_{\gamma})=q(\negr{\beta})\, q(b)\, q(\sigma^2_{\gamma})\,\prod_{i=1}^K q({\gamma}_i)$. Let
\begin{eqnarray}
\textit{diff}_{\negr{\beta}}=\E_{q}[\log p(\negr{\beta})]-\E_{q}[\log q(\negr{\beta})] \nonumber
\end{eqnarray}
\begin{eqnarray}
\textit{diff}_{\negr{\gamma}}=\E_{q}\Big[\sum_{i=1}^K \log p({\gamma}_i\, \vert \sigma^2_{\gamma})\Big]-\E_{q}\Big[\sum_{i=1}^K \log q({\gamma}_i)\Big] \nonumber
\end{eqnarray}
\begin{eqnarray}
\textit{diff}_{b}=\E_{q}[\log p(b)]-\E_{q}[\log q(b)] \nonumber
\end{eqnarray}
\begin{eqnarray}
\textit{diff}_{\sigma^2_{\gamma}}=\E_{q}[\log p(\sigma^2_{\gamma})]-\E_{q}[\log q(\sigma^2_{\gamma})] \nonumber
\end{eqnarray}
Then we can calculate the ELBO as follows with proof details provided in the Appendix B.
\begin{eqnarray}
\text{ELBO}(q)=\E_{q}[\log p(\vect{D}\,\vert\,\negr{\beta}, \negr{\gamma}, b)]+\textit{diff}_{\negr{\beta}}+\textit{diff}_{\negr{\gamma}}+\textit{diff}_{b}+\textit{diff}_{\sigma^2_{\gamma}} \label{VBAFT.frailty.eq:ELBO}
\end{eqnarray}
where 
\begin{eqnarray}
\E_{q}[\log p(\vect{D}\,\vert\,\negr{\beta}, \negr{\gamma}, b)]\addeq -\delta\,\E_{q(b)}(\log b)+ \E_{q(b)}\big(\frac{1}{b}\big) \sum_{i=1}^K\sum_{j=1}^{n_i}\Big(\delta_{ij}-(1+\delta_{ij})\varphi_{ij}\Big)\Big(y_{ij}-\vect{X}_{ij}^T\E_{q(\negr{\beta})}(\negr{\beta})-\E_{q({\gamma}_i)}{\gamma}_i\Big)\nonumber
\end{eqnarray}
\begin{eqnarray}
\textit{diff}_{\negr{\beta}} \addeq -\frac{1}{2}v_0[\text{trace}(\Sigma^*)+(\negr{\mu}^*-\negr{\mu}_0)^T(\negr{\mu}^*-\negr{\mu}_0)] + \frac{1}{2}\log (\vert\Sigma^*\vert), \nonumber
\end{eqnarray}
\begin{eqnarray}
\textit{diff}_{\negr{\gamma}} \addeq-\frac{K}{2}\E_{q(\sigma^2_\gamma)}(\log \sigma^2_\gamma)-\frac{1}{2}\E_{q(\sigma^2_\gamma)}\big(\frac{1}{\sigma^2_\gamma}\big)\sum_{i=1}^K \E_{q(\gamma_i)}(\gamma_i^2) -\frac{1}{2} \sum_{i=1}^K\big(\log \sigma_i^{2*}\big),\nonumber
\end{eqnarray}
\begin{eqnarray}
\textit{diff}_{b} \addeq (\alpha^*-\alpha_0)\E_{q(b)}(\log b) + (\omega^*-\omega_0)\E_{q(b)}\big(\frac{1}{b}\big) - \alpha^*\log \omega^*,\; \text{and} \nonumber
\end{eqnarray}
\begin{eqnarray}
\textit{diff}_{\sigma^2_{\gamma}} \addeq (\lambda^*-\lambda_0)\E_{q(\sigma^2_{\gamma})}(\log \sigma^2_{\gamma}) + (\eta^*-\eta_0)\E_{q(\sigma^2_{\gamma})}\big(\frac{1}{\sigma^2_{\gamma}}\big) - \lambda^*\log \eta^* \nonumber
\end{eqnarray}

We now present the calculation of the expectations in the update equations and the ELBO calculation. All the expectations are taken with respect to the approximated variational distributions. Since $q^*(b)$ is an Inverse-Gamma$(\alpha^*, \omega^*)$, we have
\begin{eqnarray}
\E_{q(b)}\Big(\frac{1}{b}\Big)=\frac{\alpha^*}{\omega^*},\nonumber
\end{eqnarray}
\begin{eqnarray}
\E_{q(b)}\Big( \frac{1}{b^2} \Big) = \E_{q(b)}\Big[\Big( \frac{1}{b} \Big)^2 \Big] =\text{Var}_{q(b)}\Big[\Big(\frac{1}{b}\Big)\Big]+\Big[\E_{q(b)}\Big(\frac{1}{b}\Big)\Big]^2 = \frac{\alpha^*}{\omega^{*2}}+\frac{\alpha^{*2}}{\omega^{*2}}= \frac{\alpha^*+\alpha^{*2}}{\omega^{*2}}, \; \text{and} \nonumber
\end{eqnarray}
\begin{eqnarray}
\E_{q(b)}(\log b)=\log(\omega^*)-\psi(\alpha^*), \nonumber
\end{eqnarray}
where $\psi$ is the digamma function defined as $\psi(x)=\frac{d}{dx}\log \Gamma(x)$.

Similarly, 
\begin{eqnarray}
\E_{q(\sigma^2_{\gamma})}\Big(\frac{1}{\sigma^2_{\gamma}}\Big)=\frac{\lambda^*}{\eta^*}\quad \text{and} \quad
\E_{q(\sigma^2_{\gamma})}(\log \sigma^2_{\gamma})=\log(\eta^*)-\psi(\lambda^*). \nonumber
\end{eqnarray}


\begin{algorithm}[!ht]
  \textbf{Data:} {a sample of independent log observed time $y_{ij}$, their corresponding covariate vectors $\vect{X}_{ij}$ and the right censoring indicator $\delta_{ij}, i=1, ...,K, j=1, ..., n_i$ for the $j^{th}$ observation from the $i^{th}$ group; values of hyperparameters: $\negr\mu_0, v_0$, $\alpha_0$, $\omega_0$, $\lambda_0$ and $\eta_0$; convergence threshold $\Delta$ and maximum number of iterations $M$} \\
  \textbf{Result:} {posterior distributions of $\negr\beta$, ${\gamma}_i, i=1,..., K$, $b$ and $\sigma^2_{\gamma}$, and their parameters: $\Sigma, \negr\mu, \sigma^2_i, {\tau}_i, \alpha, \omega, \lambda, \eta$} \\
  \textbf{Initialization}: initialize $\omega=\omega_0$, $\negr\mu=\negr\mu_0$, ${\tau}_i=0$ and $\eta = \eta_0$, set $m=0$ and $\text{ELBO}=0$ \\
  \textbf{Calculation}: obtain $\alpha$ by $\alpha = \alpha_0 + \delta$ with $\delta=\sum_{i=1}^K\sum_{j=1}^{n_i}\delta_{ij}$ and $\lambda$ by $\lambda=\lambda_0 + K/2$\\
  \textbf{while} \textit{iteration $m<M$ and difference of ELBO $>\Delta$}\quad \textbf{do:}\\
  \begin{tabular}{ @{\hspace{\tabcolsep}} | l }
   $m=m+1$;\\
    $\Sigma^{(m)} \leftarrow \bigg[v_0\textbf{I}_{p \times p}+2\E_{q(b)}\big(\frac{1}{b^2}\big) \sum_{i=1}^K\sum_{j=1}^{n_i}(1+\delta_{ij})\zeta_{ij}\vect{X}_{ij}\vect{X}_{ij}^T\bigg]^{-1}$; \\
    $\negr\mu^{(m)}\leftarrow \Bigg[\bigg\{v_0\,\negr{\mu}_0^T + \sum_{i=1}^K\sum_{j=1}^{n_i}\bigg( \Big(\E_{q(b)}\big(\frac{1}{b}\big)\Big(-\delta_{ij}+(1+\delta_{ij})\rho_{ij}\Big) +2\E_{q(b)}\big(\frac{1}{b^2}\big) (1+\delta_{ij})\zeta_{ij}\big(y_{ij}-\E_{q({\gamma}_i)}({\gamma}_i)\big) \Big)\vect{X}_{ij}^T\bigg)\bigg\}\,\Sigma^{(m)}\Bigg]^{T}$;\\
    $\sigma_{i}^{2(m)} \leftarrow \big[\E_{q(\sigma^2_{\gamma})}\big(\frac{1}{\sigma^2_{\gamma}}\big)+2\E_{q(b)}\big(\frac{1}{b^2}\big)\sum_{j=1}^{n_i}(1+\delta_{ij})\zeta_{ij}\big]^{-1}, i = 1, ..., K$; \\
    ${\tau}_i^{(m)} \leftarrow \sigma_{i}^{2(m)} \sum_{j=1}^{n_i}\Big[\E_{q(b)}(\frac{1}{b})\big(-\delta_{ij}+(1+\delta_{ij})\rho_{ij}\big) +2\E_{q(b)}(\frac{1}{b^2})(1+\delta_{ij})\zeta_{ij}\big(y_{ij}-\vect{X}_{ij}^T\E_{q(\negr{\beta})}\negr{\beta}\big)\Big], i=1, ..., K$; \\
      $\omega^{(m)} \leftarrow \omega_0-\sum_{i=1}^K\sum_{j=1}^{n_i}\Big(\delta_{ij}-(1+\delta_{ij})\varphi_{ij}\Big)\Big(y_{ij}-\vect{X}_{ij}^T\E_{q(\negr{\beta})}(\negr{\beta})-\E_{q(\gamma_i)}\gamma_i\Big)$; \\
      $\eta^{(m)} \leftarrow \eta_0 + \frac{1}{2}\sum_{i=1}^K\E_{q(\gamma_i)}\gamma_i^2$;\\
      calculate the current ELBO, $\text{ELBO}^{(m)}$; \\
      calculate the difference of ELBO $=\text{ELBO}^{(m)}-\text{ELBO}^{(m-1)}$;\\
  \end{tabular} \\ 
\textbf{end}
  \caption{Variational Bayesian inference of correlated survival data using a shared frailty log-logistic AFT model}
  \label{VBsurvival:frailty}
\end{algorithm}
\vspace{-1cm}
\section{Simulation study}\label{sec:simulation}
\subsection{Design of simulation}
We conduct a simulation study to evaluate the performance of our proposed VB algorithms across different scenarios by varying the number of clusters and the number of observations within each cluster.

We generate the log of survival time for the $j^{th}$ subject in the $i^{th}$ cluster, $\log(T_{ij})$, $i = 1, ..., K$ and $j = 1, ..., n_i$ as follows:
$$\log T_{ij}=0.5+\beta_1 x_{ij1}+ \beta_2 x_{ij2} +\gamma_{i} +b\epsilon_{ij},$$
where $x_{ij1}$, $x_{ij2}$, and $\epsilon_{ij}$ are mutually independently generated with $x_{ij1}\sim N(1, 0.2^2)$, $x_{ij2}\sim \text{Bernoulli}(0.5)$ and $\epsilon_{ij} \sim \text{logistic}(0,1)$. The values of $\beta_1$, $\beta_2$ and $b$ are chosen as 0.2, 0.8 and 0.8, respectively. The random intercept for the $i^{th}$ cluster, $\gamma_{i}$, is generated from $N(0, \sigma_{\gamma}^2)$ with $\sigma_{\gamma}^2=1$. The censoring time for the $j^{th}$ subject in the $i^{th}$ cluster, $C_{ij}$, is generated from a uniform distribution, $\text{uniform}(0, d)$, where $d$ is a positive value controlling the percentage of censoring. Then $t_{ij} = \min(T_{ij}, C_{ij})$ and $\delta_{ij} = \mathbbm{1}(T_{ij} \leq C_{ij})$. Take $d = 48$ to achieve a 15\% censoring rate in our simulations. Investigation of the effect of censoring rates can be found in Xian et al. \citep{Xian_2024} where they show that as the censoring rate increases, an increase in mean squared error (MSE) of estimating parameters in an AFT model using a VB algorithm was observed.

We explore various scenarios by varying the number of clusters, $K$, and the number of observations within each cluster, $n_i=n$ for all $i=1, ..., K$. Across the experiments, we consider $K$ values from the set $\{15, 30, 50, 80\}$ and $n$ values from $\{5, 15, 30, 50\}$, resulting in a total of 16 unique scenarios. Our objective is to assess the estimation performance of the VB algorithm concerning the variations in $K$ and $n$. We consider a prior setting with $\negr{\mu}_0 = (0, 0, 0)^T$, $v_0=0.1$, $\alpha_0=\lambda_0=3$, and $\omega_0=\eta_0=2$, which indicates no strong prior information on the parameters. The ELBO convergence threshold is set as $0.01$ which is the default recommendation \cite{Yao_2018}, and the maximum number of iterations is $100$.

We conduct $N = 500$ runs (replicates) in each considered scenario and apply our proposed VB algorithm to derive the approximated posterior distribution of each parameter to each run. The mean of each approximated posterior distribution is used as the parameter estimate for the corresponding parameter. The empirical bias and sample standard deviation (SD) as well as the empirical MSE for each estimate are obtained, where
\begin{eqnarray}
\text{MSE}=\frac{\sum_{i=1}^N (\theta_0 -\hat{\theta}_i)^2}{N} \nonumber
\end{eqnarray}
and $\hat{\theta}_i$ is the estimate of parameter $\theta$ in the $i^{th}$ replicate, and $\theta_0$ is the true value. In addition, for each parameter of interest, we report the empirical 95\% coverage rate (CR) calculated as 
$$\text{CR}=\frac{\sum_{i=1}^N I_i}{N}
$$
where $I$ is the indicator variable which takes $1$ if the true parameter value $\theta_0$ falls into the 95\% credible interval.

\subsection{Simulation results}
\begin{figure}[!ht]
\centering
\begin{subfigure}{.5\textwidth}
  \centering
  \includegraphics[width = 8.5cm]{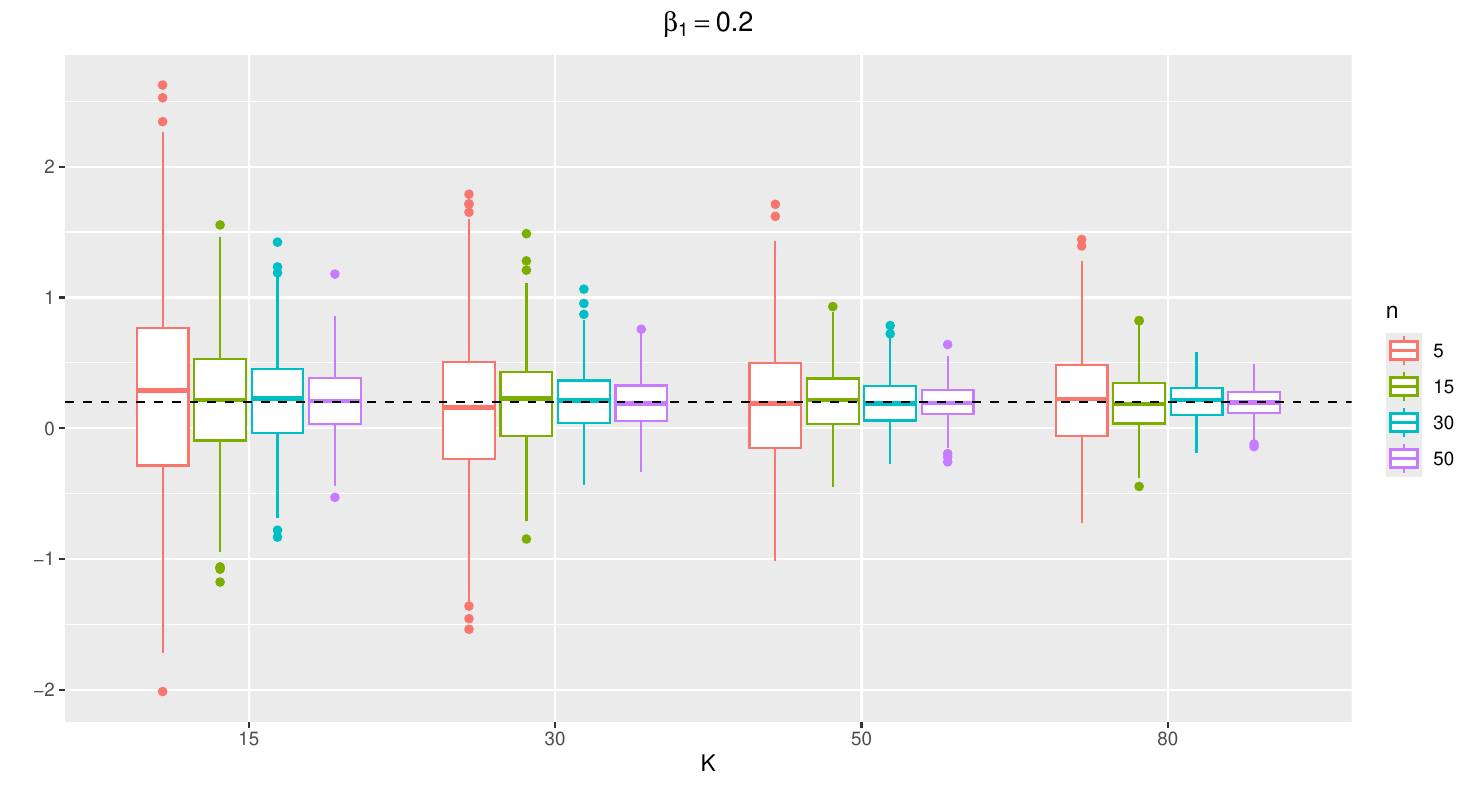}
\end{subfigure}%
\begin{subfigure}{.5\textwidth}
  \centering
  \includegraphics[width = 8.5cm]{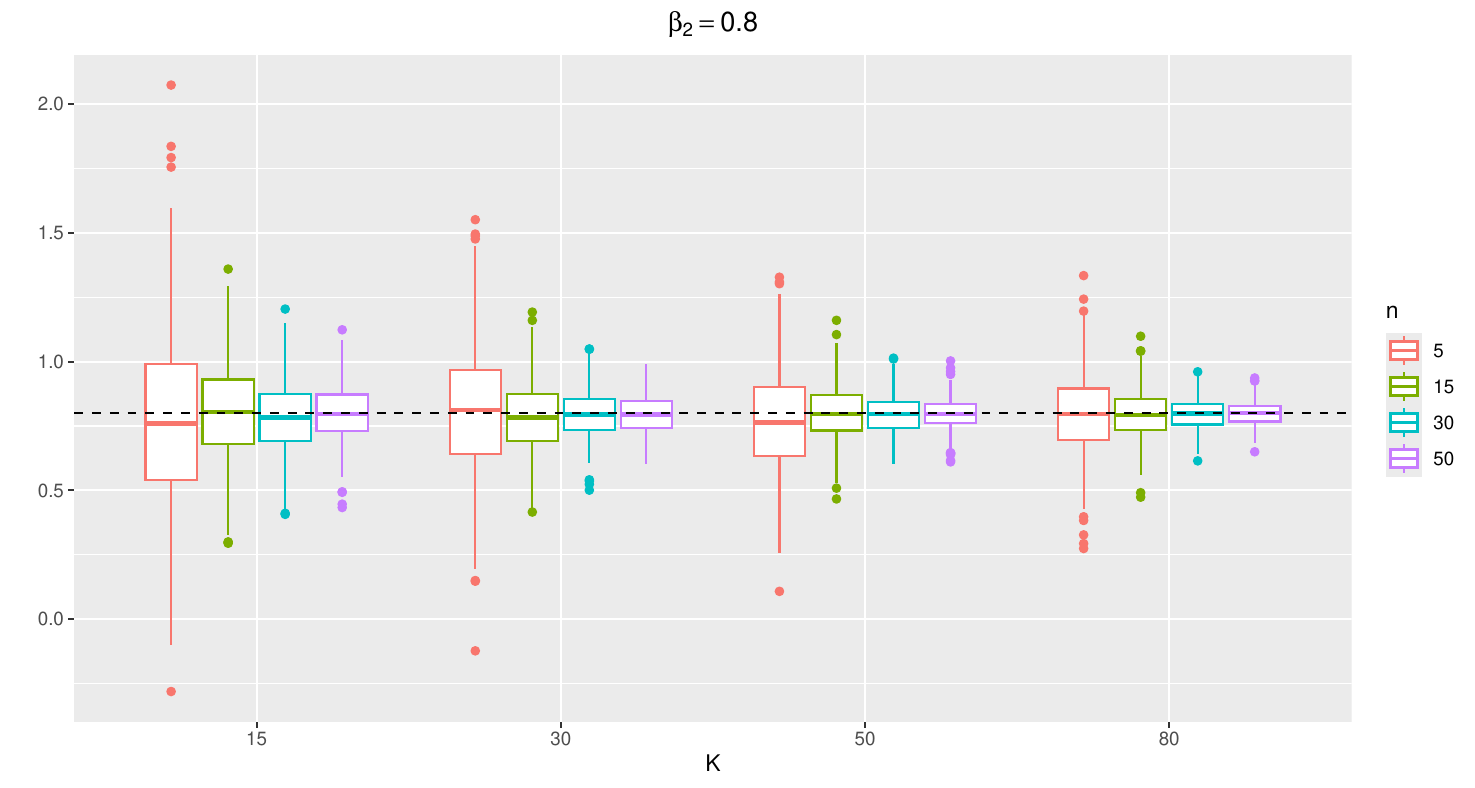}
\end{subfigure} 
\begin{subfigure}{.5\textwidth}
  \centering
  \includegraphics[width = 8.5cm]{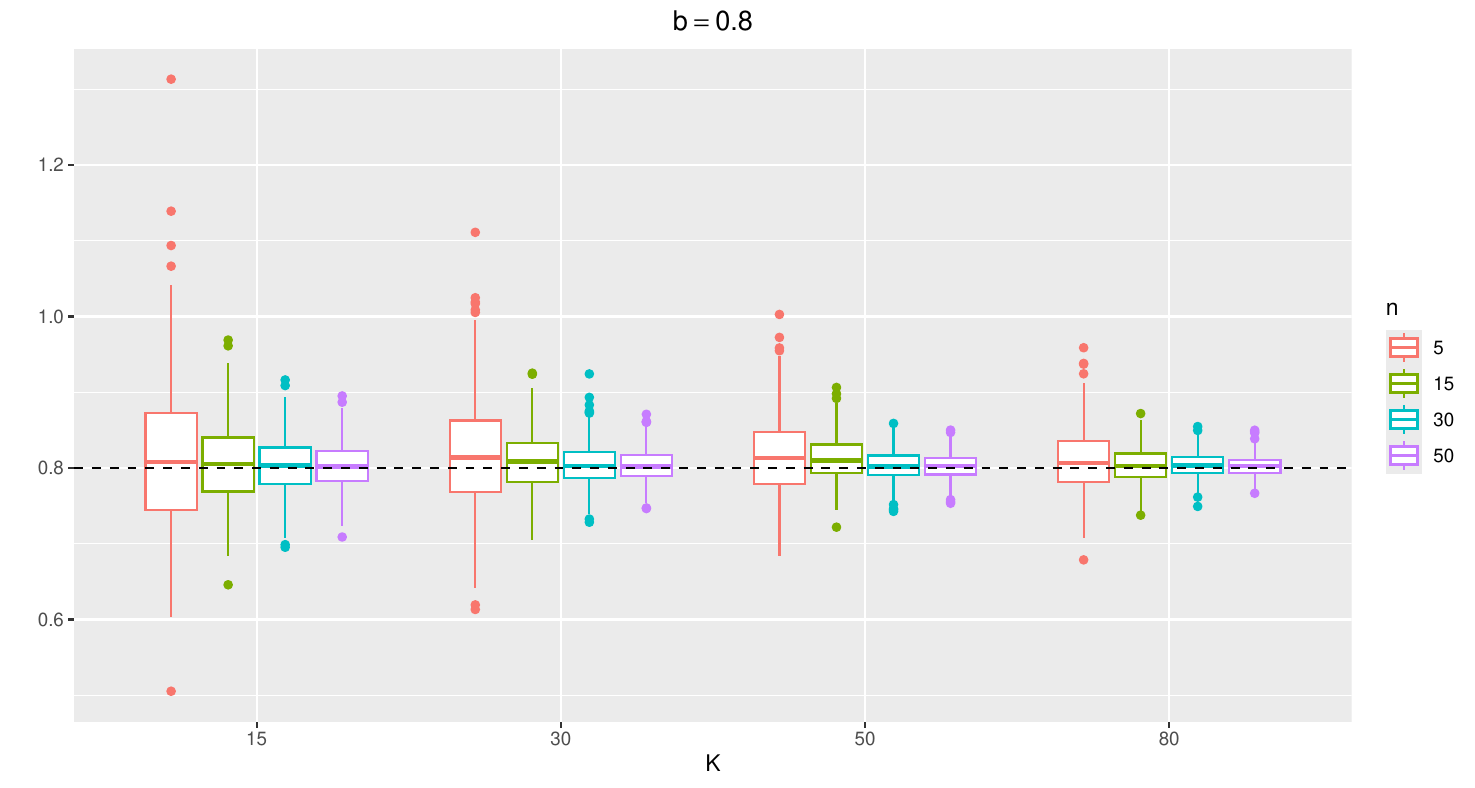}
\end{subfigure}%
\begin{subfigure}{.5\textwidth}
  \centering
  \includegraphics[width = 8.5cm]{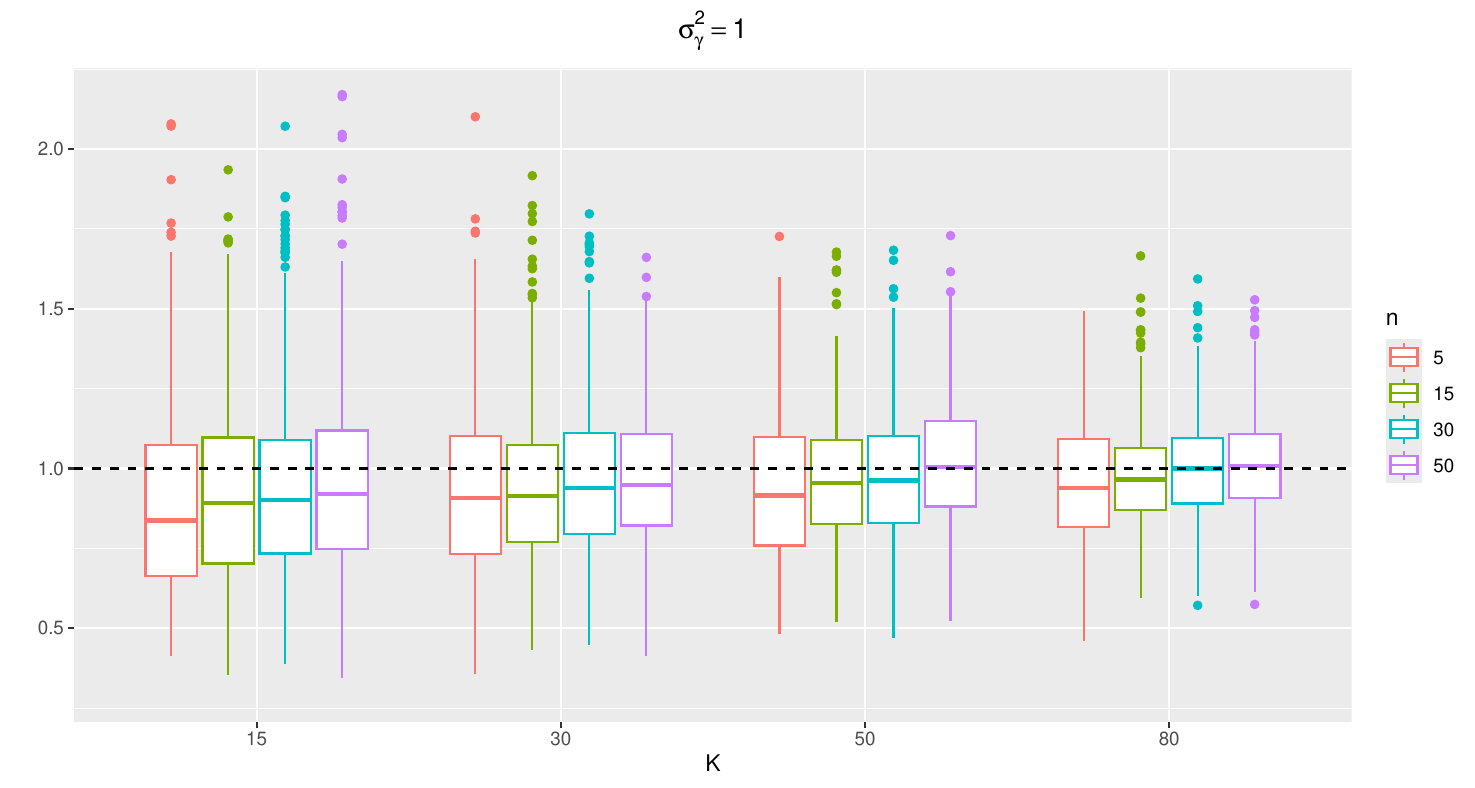}
\end{subfigure}%
\caption{\hspace{0.1cm} Boxplots of parameter estimates using posterior means from 500 replicates under various scenarios with different number of clusters $K$ and cluster sizes $n$ based on our proposed VB algorithm. The horizontal dashed line on each plot represents the true value of the corresponding parameter used when generating the data.}
\label{Sim.one.res.boxplots}
\end{figure}

The empirical bias, sample SD, empirical MSE and empirical CR of estimating $\beta_1$, $\beta_2$, $b$, and $\sigma_{\gamma}^2$ from 500 replicates are summarized in Table \ref{sim.one.VB}. Considering $\beta_1$, $\beta_2$, and $b$, we observe that when the number of clusters $K$ is fixed, increasing the cluster size $n$ from 5 to 50 does not always significantly affect the empirical bias. However, there is a noticeable decrease in the sample SD, leading to a pronounced reduction in the empirical MSE. A similar trend is observed when the cluster size is fixed while increasing $K$. Figure \ref{Sim.one.res.boxplots} visually illustrates this asymptotic property through boxplots. Our primary focus lies in understanding the impact of $K$ and $n$ on estimating the variance of the random intercept, denoted as $\sigma_\gamma^2$. In most of the scenarios, when the number of clusters $K$ is fixed, we observe that the bias of estimating $\sigma_\gamma^2$ decreases as the cluster size increases, while the sample SD remains relatively stable. However, when $K$ is increased, both the empirical bias and the sample SD decrease noticeably, except for the case when $n=30$ as $K$ grows from 15 to 30. Particularly noteworthy is the scenario with $K = 80$ and $n = 50$, where we achieve a less biased and less variable estimation of the variance of the random intercept. In addition, the empirical CRs corresponding to a 95\% credible interval for each parameter across all scenarios are close to the nominal level of 95\%, with a mean of 94.1\% and a standard deviation of 0.014. Furthermore, based on a nominal level of 95\% with 500 replicates, the coverage with two standard deviations is $95\% \pm 2\text{SD}$ = $[93.1\%, 97.0\%]$ since $\text{SD}=\sqrt{0.95 \times 0.05 / 500} \approx 0.0097$. We observe that the majority of our empirical CRs are within the range of $[93.1\%, 97.0\%]$.

\begin{figure}[!ht]
\centering
\begin{subfigure}{.5\textwidth}
  \centering
  \includegraphics[height = 5.5cm]{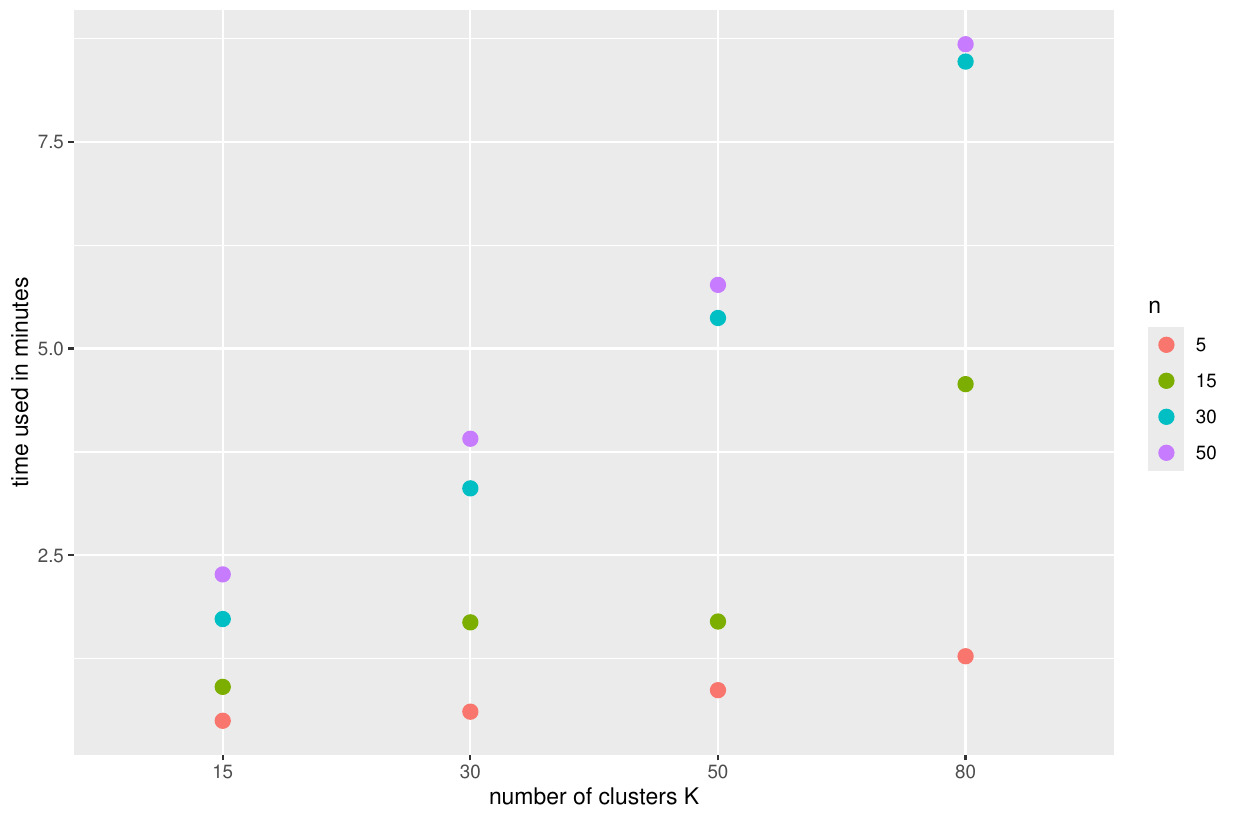}
\end{subfigure}%
\begin{subfigure}{.5\textwidth}
  \centering
  \includegraphics[height = 5.5cm]{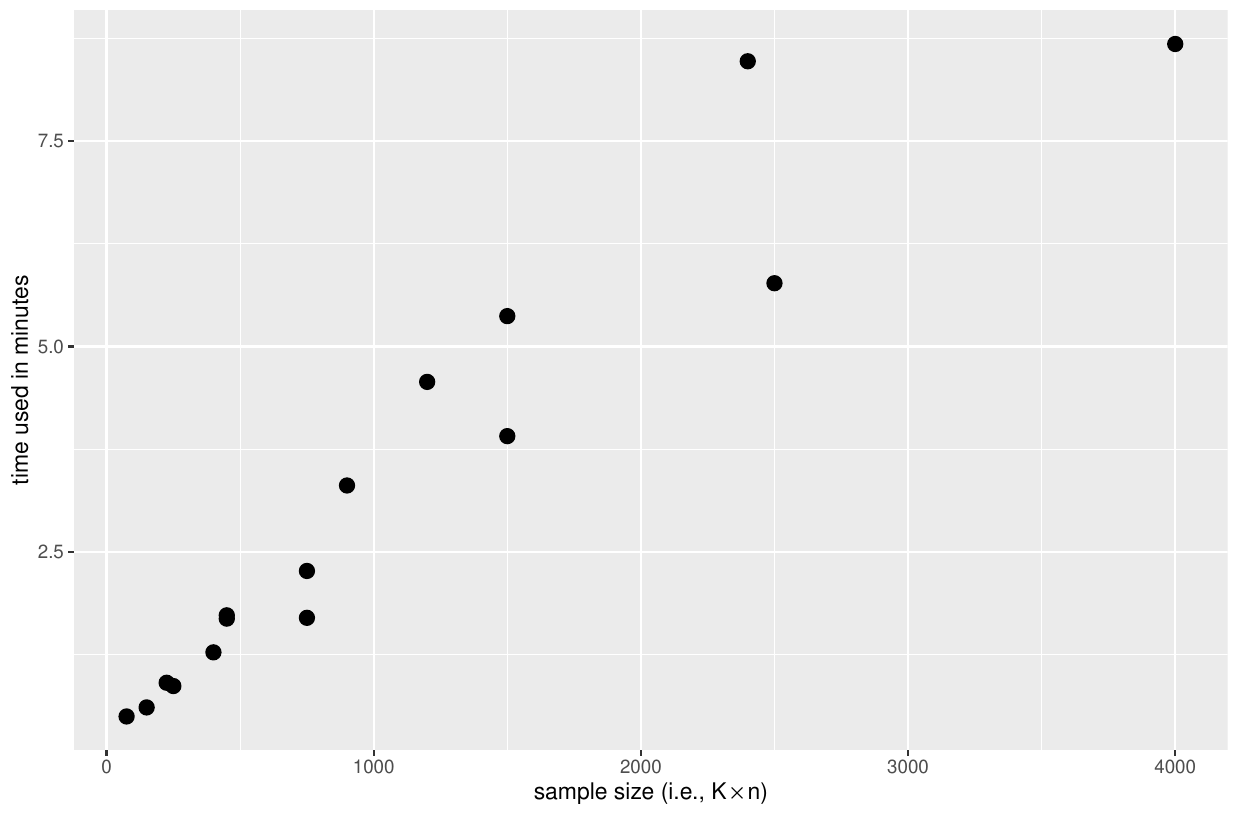}
\end{subfigure} 
\caption{\hspace{0.1cm} \textbf{Left}: the run time in minutes used for 500 replicates under various scenarios with different number of clusters $K$ and cluster sizes $n$ based on our proposed VB algorithm. \textbf{Right}: the run time in minutes used for 500 replicates in different sample sizes based on VB.}
\label{Sim.one.res.cost}
\end{figure}

The left part of Figure \ref{Sim.one.res.cost} presents the computational run time of 500 replicates in minutes against the number of clusters ($K$) with different colors for different cluster sizes $n$. Fixing $K$, as $n$ increases, the used time increases. The scatter plot in the right of Figure \ref{Sim.one.res.cost} shows the run time against the sample size ($K \times n$). As anticipated, the computational cost rises with increasing sample size.

\begin{figure}[!ht]
\centering
\begin{subfigure}{.33\textwidth}
  \centering
  \includegraphics[height = 5cm]{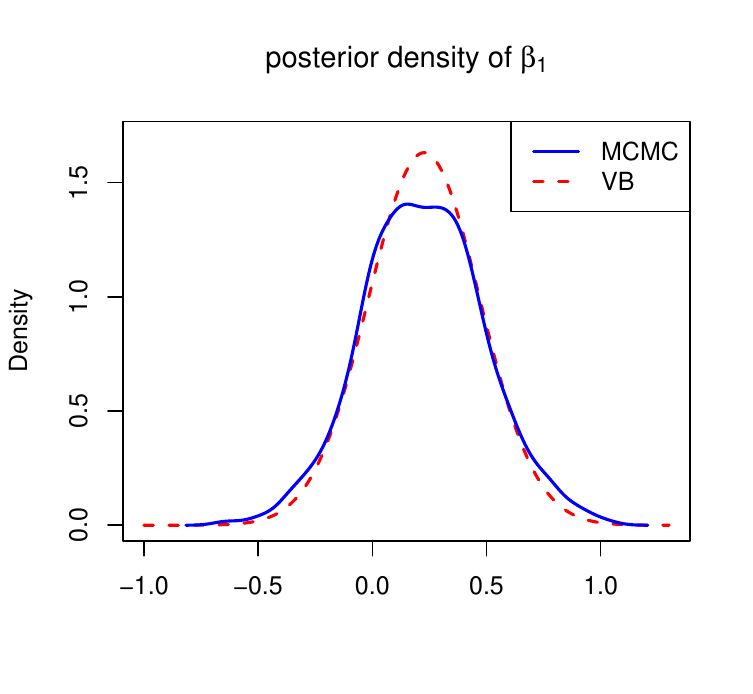}
\end{subfigure}%
\begin{subfigure}{.33\textwidth}
  \centering
  \includegraphics[height = 5cm]{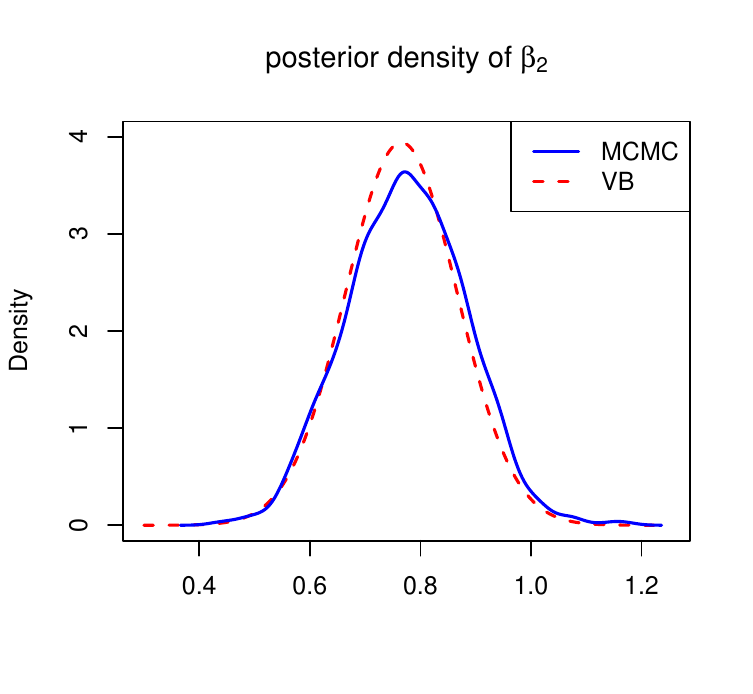}
\end{subfigure}%
\begin{subfigure}{.33\textwidth}
  \centering
  \includegraphics[height = 5cm]{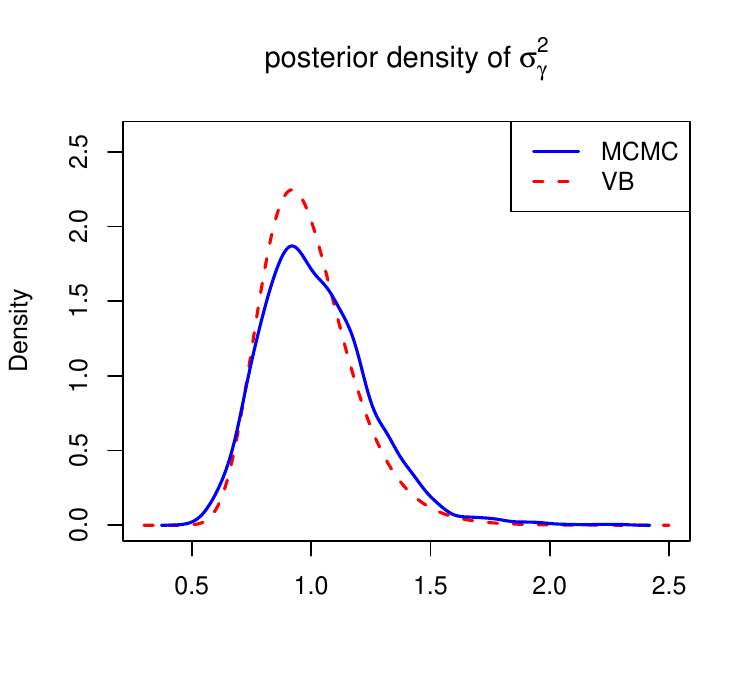}
\end{subfigure}%
\caption{\hspace{0.1cm} A comparison of posterior densities of $\beta_1$, $\beta_2$ and $\sigma_\gamma^2$ obtained from our VB method \textit{survregVBfrailty} and MCMC-based \textit{survregbayes} from a simulated data set in the scenario with $K = 50$ and $n = 15$.}
\label{MCMCVB.density_Comp}
\end{figure}

Furthermore, among those considered 16 scenarios, we select three scenarios ($K = 30$ with $n =5$, $K = 50$ with $n =15$, and $K = 80$ with $n =30$) and conduct a comparative analysis of estimation results obtained from the proposed VB algorithm with those from two alternative methods which we introduced in Section \ref{sec1}: the h-likelihood method proposed by Do Ha et al. \cite{Do_2017} and the MCMC-based \textit{survregbayes} developed by Zhou et al. \cite{Zhou_2017}. Summaries from the MCMC-based \textit{survregbayes} algorithm are derived from one Markov chain, subsampled every 5 iterates to achieve a final chain size of 2,000 after a burn-in period of 5,000 iterates \citep{Zhou_2017}. We further refer to our VB method and the h-likelihood method as \textit{survregVBfrailty} and \textit{survregHL}, respectively. Numerical estimation results are presented in Table \ref{sim.res.one}, where we focus on the comparison of estimation for $\beta_1$, $\beta_2$ and $\sigma_\gamma^2$ since the MCMC-based \textit{survregbayes} does not directly return the estimate for $b$. In the scenario with $K=30$ and $n=5$, the proposed \textit{survregVBfrailty} exhibits a larger empirical bias compared to both the \textit{survregHL} and \textit{survregbayes} methods. However, as $K$ and $n$ increase, the discrepancy in empirical bias among the three methods becomes negligible. In terms of the sample SD, on average, across the three parameters $\beta_1$, $\beta_2$, and $\sigma_{\gamma}^2$, \textit{survregVBfrailty} yields a 16.5\% smaller sample SD when $K=30$ and $n=5$, a 7.6\% smaller sample SD when $K=50$ and $n=15$, and a 3.7\% smaller sample SD when $K=80$ and $n=30$, compared to \textit{survregHL}. Compared with the \textit{survregBayes} algorithm, we observe a decrease in sample SD of 7.8\% in the $K=30$, $n=5$ scenario, mainly due to $\beta_1$, and only 2.0\% in the $K=50$, $n=15$ scenario. However, in the scenario with $K=80$, $n=30$, no average difference in sample SD is observed between VB and MCMC. Therefore, the \textit{survregVBfrailty} algorithm generally yields a smaller SD, resulting in a lower MSE compared to the other two methods in most cases. On average, across the three parameters $\beta_1$, $\beta_2$, and $\sigma_{\gamma}^2$, \textit{survregVBfrailty} achieves a 27.2\% lower MSE when $K=30$, $n=5$, a 12.0\% lower MSE when $K=50$, $n=15$, and a 13.4\% lower MSE when $K=80$, $n=30$, compared to \textit{survregHL}. Compared with \textit{survregBayes}, we observe a reduction in MSE of 12.4\% in the $K=30$, $n=5$ scenario, mainly because of $\beta_1$, and only 3.0\% in the $K=50$, $n=15$ scenario. However, in the scenario with $K=80$, $n=30$, no difference in MSE is observed between VB and MCMC. In addition, we compare the posterior densities for $\beta_1$, $\beta_2$, and $\sigma_\gamma^2$ obtained from both \textit{survregVBfrailty} and the MCMC-based \textit{survregBayes} using a simulated dataset with $K=50$ and $n=15$. These densities present a strong consistency between the VB and MCMC algorithms as illustrated in Figure \ref{MCMCVB.density_Comp}, which is also reflected in the results presented in Table \ref{sim.res.one}. 

To assess the computational efficiency of the proposed VB algorithm, we compare the run times in minutes for 500 replicates in these three scenario across \textit{survregVBfrailty}, \textit{survregHL}, and \textit{survregbayes} methods and present the results in Table \ref{time_comp}. In Table \ref{time_comp}, we observe that in the scenario with a small sample size ($K=30$ and $n=5$), there is minimal difference in the run times between the VB and h-likelihood methods. However, as both $K$ and $n$ increase, the h-likelihood method requires progressively longer computation times to obtain estimates. Particularly in the scenario with $K = 80$ and $n=30$, the computational time is over 10 times longer than that required by the VB algorithm. Comparatively, the VB algorithm demonstrates significantly higher computational efficiency when contrasted with the MCMC-based \textit{survregbayes} algorithm. Notably, in the scenario with $K = 80$ and $n=30$, the \textit{survregVBfrailty} algorithm runs approximately 150 times faster than the \textit{survregbayes} algorithm. All algorithms were implemented within R version 4.3.2 and simulations were conducted on a computer operating the Mac OS X platform, equipped with a 4.05 GHz CPU and 8 GB RAM.

\begin{landscape}
\vspace*{\fill}
\begin{table}[!ht]
\begin{center}
\caption{\hspace{0.1cm} Numerical estimation results (point estimate using the mean of the corresponding posterior distribution) including the empirical Bias, sample SD, empirical MSE and coverage rate (CR) for parameters in each scenario with different number of clusters $K$ and cluster sizes $n$ from the proposed VB algorithm.}
\label{sim.one.VB}
\begin{tabular}{llllllllllllllllll}
\toprule
\multicolumn{2}{c}{} & \multicolumn{4}{c}{$\beta_1$} &  \multicolumn{4}{c}{$\beta_2$} & \multicolumn{4}{c}{$b$}& \multicolumn{4}{c}{$\sigma_\gamma^2$} \\
\cmidrule(rl){3-6} \cmidrule(rl){7-10} \cmidrule(rl){11-14}\cmidrule(rl){15-18}
$K$  & $n$        &  Bias       & SD       & MSE &CR     & Bias     & SD      & MSE &CR & Bias     & SD      & MSE &CR &  Bias     & SD      & MSE  &CR  \\ \midrule
\multicolumn{1}{c}{\multirow{4}{*}{15}} & 5 & 0.066	&0.764&	0.587 &94.2	&	-0.026&	0.352&	0.124& 90.0&		0.005&	0.095&	0.009& 94.5&	-0.101&	0.312	&0.107 &91.4\\
\multicolumn{1}{c}{}                          &       15&	0.018&	0.489&	0.239 &93.2	&	0.007	&0.192&	0.037& 94.0&	0.006&	0.052&	0.003&94.8&	-0.081&	0.287&	0.089 &93.1                \\
\multicolumn{1}{c}{}                          &       30&	0.006&	0.354&	0.125&93.2&	-0.017&	0.139&	0.019&92.4&	-0.006&	0.036&	0.001&96.0&	0.003&	0.290	&0.087 &95.5                \\
\multicolumn{1}{c}{}            &       50&0.009	&0.254	&0.065&95.6	&	-0.001&	0.104&	0.011&94.8&	0.003&	0.029&	0.001&94.7&	-0.040&	0.299&	0.100&94.2  \\\midrule

\multicolumn{1}{c}{\multirow{4}{*}{30}} & 5 & 0.004&	0.589	&0.346&92.8	&	-0.015&	0.257&	0.066&92.8&	0.015&	0.071&	0.005&94.4&	-0.092&	0.274 & 0.084&92.6\\
\multicolumn{1}{c}{}                          &       15&0.002	&0.360	&0.129	&93.0&	-0.018	&0.140&	0.020&93.0	&	0.008	&0.038&	0.001	&96.1&	-0.057	&0.242	&0.062&91.1	  \\
\multicolumn{1}{c}{}                          &       30&	0.012&	0.239	&0.057 &94.6	&	-0.004	&0.088&	0.008&95.7	&	0.005	&0.026	&0.001&95.6	&	-0.036	&0.248&	0.063 &92.4                \\
\multicolumn{1}{c}{}            &       50&-0.007&	0.192&	0.037&93.0	&	-0.005&	0.074&	0.005&95.0	&0.003&	0.020&	$< 0.001$&95.8	&	-0.026	&0.228	&0.052&95.2  \\\midrule

\multicolumn{1}{c}{\multirow{4}{*}{50}} & 5 & -0.011&	0.470	&0.220	&94.3&	-0.033&	0.199&	0.041&91.4	&	0.014&	0.051	&0.003&95.1	&	-0.062&	0.235&	0.059&93.2\\
\multicolumn{1}{c}{}                          &       15&	0.003&	0.253&	0.064&95.2	&	-0.001&	0.104&	0.011&94.8	&	0.018&	0.029&	0.001	&94.2&	-0.033&	0.203	&0.042&92.4                 \\
\multicolumn{1}{c}{}                          &       30&	-0.006&	0.192&	0.037&92.3&		-0.006&	0.074&	0.006&94.4&	0.003&	0.020&	$< 0.001$	&95.6&	-0.027&	0.191	&0.037 &93.3                \\
\multicolumn{1}{c}{}            &       50&-0.002&	0.143&	0.020	&94.4&	-0.004	&0.057	&0.003&95.6	&	0.002&	0.016&	$< 0.001$&95.7	&	0.016&	0.201&	0.040&93.4 \\\midrule

\multicolumn{1}{c}{\multirow{4}{*}{80}} & 5 & 0.013&	0.384&	0.147&93.2	&-0.005	&0.161&	0.026&91.3	&	0.014&	0.041	&0.002&95.2	&-0.048&	0.197&	0.041&94.4\\
\multicolumn{1}{c}{}                          &       15&	-0.008	&0.223&	0.050	&93.4&	-0.005&	0.090&	0.008&92.4	&	0.008&	0.023&	0.001&95.5	&	-0.023&	0.160	&0.026 &92.2               \\
\multicolumn{1}{c}{}                          &       30&	0.008&	0.141&	0.020&95.8	&	-0.005&	0.058	&0.003&95.1	&	-0.003&	0.015 &	$< 0.001$&95.6	&	0.003	&0.161	&0.026 &92.8              \\
\multicolumn{1}{c}{}            &       50&-0.005	 &0.108	 &0.012&95.4	 &	-0.002	 &0.043 &	0.002&95.5	 &	0.003	 &0.013	 &$< 0.001$	&95.5 &	0.014	 &0.157	 &0.025&94.4\\
\bottomrule
\end{tabular}
\end{center}
\end{table}
\vspace*{\fill}
\end{landscape}
\begin{table}[ht]
\begin{center}
\caption{\hspace{0.1cm} A comparison of numerical estimation results including the empirical Bias, sample SD and MSE, from our VB method \textit{survregVBfrailty}, the h-likelihood method \textit{survregHL} and MCMC-based \textit{survregbayes} method in each scenario ($K$: number of clusters, $n$: number of observations in each cluster).}
\label{sim.res.one}
\begin{tabular}{ccccccccccc}
\toprule
\multicolumn{2}{c}{} & \multicolumn{3}{c}{\textit{survregVBfrailty}} &  \multicolumn{3}{c}{\textit{survregHL}} & \multicolumn{3}{c}{\textit{survregbayes}} \\
\cmidrule(rl){3-5} \cmidrule(rl){6-8} \cmidrule(rl){9-11}
Scenario  & Parameter  & Bias       & SD       & MSE & Bias     & SD      & MSE & Bias     & SD      & MSE   \\ \midrule
\multicolumn{1}{c}{\multirow{3}{*}{$K = 30$, $n = 5$}} & $\beta_1$     & 0.004	&0.589	&0.346& 0.001&	0.664&	0.440& 0.001&	0.664&	0.440\\
\multicolumn{1}{c}{} & $\beta_2$     & -0.015&	0.257&	0.066&	-0.002&	0.270
&0.073 &-0.002	&0.266&	0.071\\
\multicolumn{1}{c}{}  & $\sigma_{\gamma}^2$ & -0.092&	0.274	&0.084	&	0.001&	0.412&	0.170&		-0.042&	0.300&	0.092 \\ \midrule
\multicolumn{1}{c}{\multirow{3}{*}{$K = 50$, $n = 15$}} & $\beta_1$     & 0.003	&0.253	&0.064	&	0.001&	0.268&	0.072&		-0.005	&0.259&	0.067 \\
\multicolumn{1}{c}{} & $\beta_2$     & -0.001&	0.104&	0.011	&	0.007&	0.107&	0.011&		-0.004&	0.104&	0.011\\
\multicolumn{1}{c}{}  & $\sigma_{\gamma}^2$ & -0.033&	0.203&	0.042&	0.011	&0.237	&0.056		&-0.001&	0.211&	0.044 \\\midrule
\multicolumn{1}{c}{\multirow{3}{*}{$K = 80$, $n = 30$}} & $\beta_1$     & 0.008&	0.141&	0.020	&	0.009&	0.143&	0.021&		-0.009&	0.142&	0.020 \\
\multicolumn{1}{c}{} & $\beta_2$     & -0.005&	0.058&	0.003&	0.001&	0.060&	0.004	&	0.001	&0.058	&0.003\\
\multicolumn{1}{c}{}  & $\sigma_{\gamma}^2$ & 0.003&	0.161&	0.026	&	0.007&	0.172&	0.029&		0.001&	0.160	&0.026 \\
\bottomrule
\end{tabular}
\end{center}
\end{table}
\vspace{-1.2cm}
\begin{table}[ht]
\begin{center}
\caption{\hspace{0.1cm} Times in minutes for 500 replicates from our VB algorithm \textit{survregVBfrailty}, the h-likelihood method \textit{survregHL} and the MCMC-based \textit{survregbayes} algorithm, respectively, under each scenario.}
\label{time_comp}
\begin{tabular}{cccc}
\toprule
Scenario & \textit{survregVBfrailty} & \textit{survregHL} & \textit{survregbayes} \\ \midrule
$K = 30$, $n = 5$ & 0.61	&0.67	&47.65\\ \midrule
$K = 50$, $n = 15$ & 1.70	&9.59&	232.31\\ \midrule
$K = 80$, $n = 30$ & 8.47	&124.53&	1302.80 \\
\bottomrule
\end{tabular}
\end{center}
\end{table}
\section{Application to ventilation duration analysis}\label{sec:real}
In this section, we apply the shared frailty log-logistic AFT model with the proposed VB algorithm to the ICU data as we descried in the Introduction section and conduct a retrospective study on the ventilation duration time. We aim to investigate the ICU site-specific random effect on patient's ventilation duration. We extend the work by Kobara et al. \citep{Kobara_2023} by incorporating the uncertainty from the ICU sites via a group-specific random intercept under a Bayesian analysis framework. 


The CCIS ICU data were collected between July 2015 and December 2016 and contained 49,467 patients receiving invasive mechanical ventilation upon arrival to ICU. About 3\% of these patients were discharged/transferred to a Complex Continuing Care Facility, other hospitals, the Level 3 Unit, and Outside the ICU while still on a ventilator, and therefore, their ventilation time were considered as right-censored data \citep{Kobara_2023}. The data were from 66 ICU sites (centers) and each site has a unique site code, for example, 3970. In the CCIS dataset, the average number of ventilated patients per ICU site over the study period of about 1.5 years is 749.5. We consider the significant covariates investigated by Kobara et al. \citep{Kobara_2023} as fixed effects which are admission source (e.g., from operation rooms), admission diagnosis, patient type (medical or surgical), scheduled admission (yes or no), scheduled surgery (yes or no), referring physician specialty, other interventions (yes or no), central venous line (CVL, yes or no), arterial line (AL, yes or no), intra-cranial pressure monitor (IPM, yes or no), extracorporeal membrane oxygen (EMO, yes or no), intra-aortic balloon pump (IABP, yes or no), age group (18-39, 40-80 or above 80 years of age), pre-LOS (no more than 1 day, between 2 and 7 days, or no less than 7 days), and the MODS score (none with MODS $\leq 1$, minimal with $1-4$ scores, mild with $4-8$ scores, moderate with $8-12$ scores or, severe with scores $> 12$). 

To apply our proposed VB algorithm in real data analysis, we consider the same weak prior setting in our simulation study described in Section 4: $\negr{\mu}_0 = \textbf{0}$, $v_0=0.1$, $\alpha_0=\lambda_0=3$, and $\omega_0=\eta_0=2$. Table \ref{tab.results.real} displays the estimated regression coefficients from the fitted shared frailty log-logistic AFT model. For comparative analysis, we also employed the h-likelihood method \textit{survregHL} and the MCMC-based \textit{survregbayes} to model the data incorporating shared frailty. In the absence of shared frailty, we used the likelihood-based \textit{survreg} from the \texttt{survival} package in R, and the VB method \textit{survregVB} proposed by Xian et al. \citep{Xian_2024}. Furthermore, for the Bayesian methods (\textit{survregVBfrailty}, \textit{survregVB}, and \textit{survregbayes}), we report the 95\% credible intervals, whereas for the likelihood-based methods, we provide the 95\% confidence intervals. In Table \ref{tab.results.real}, we first observe that there is no significant difference in the estimated regression coefficients between \textit{survregVBfrailty} and \textit{survregbayes}. However, some of the estimated coefficients from \textit{survregHL}, such as scheduled surgery, CVL, AL, IPM and EMO, are different from those based on \textit{survregVBfrailty} and \textit{survregbayes}, which is further discussed in Section \ref{Sec:Dis}. In comparison to the models with frailty, the \textit{survregVB} and \textit{survreg} methods, which do not account for frailty, exhibit some differences in estimating certain coefficients. However, they maintain a strong overall consistency in estimating the regression coefficients with the results from the frailty models. This consistency is expected, as the inclusion of frailty does not affect the fixed effects, implying that the regression lines from models with or without frailty should be parallel. To illustrate the difference in interval estimation between models with and without frailty, we calculated the mean overlap percentage of the two 95\% credible intervals from VB with or without frailty. Using the 95\% credible intervals from VB without frailty (\textit{survregVB}) as references, we found a 77.3\% overlap on average. This means that if we fit a model using VB without considering the ICU site as a shared frailty, about 77\% of the credible interval will fall within the corresponding credible interval obtained from the model that includes the shared frailty.

In what follows, we summarize the fixed effects on ventilation duration based on estimates from our proposed VB algorithm, \textit{survregVBfrailty}. Regarding the admission source, patients arriving from the emergency department (ED), operating room (OR), or ward have ventilation times that are $(1 - \exp(-0.134))100\%=12.54\%$ shorter (credible interval (CI): [9.24\%, 12.72\%]), 27.89\% shorter (CI: [25.62\%, 30.02\%]), and 13.06\% shorter (CI: [9.70\%, 16.31\%]), respectively, compared to patients admitted from a downstream unit (baseline). Conversely, patients admitted from home or another hospital have ventilation durations that are 10.63\% (CI: [-10.40\%, 23.49\%]) and 10.30\% (CI: [7.04\%, 13.54\%]) longer, respectively. Since the CI for patients admitted from home includes zero, we conclude that admission from home is not a statistically significant factor. Additionally, patients from other sources, such as from outside the province, have ventilation durations that are 12.98\% longer (CI: [5.13\%, 21.29\%]).

Compared to cardiovascular patients, those with gastrointestinal, neurological, and trauma diagnoses experience significantly longer ventilation times. Specifically, ventilation duration increases by 29.69\% (CI: [25.36\%, 34.18\%]) for gastrointestinal patients, 33.11\% (CI: [29.05\%, 37.44\%]) for neurological patients, and 77.71\% (CI: [69.72\%, 86.08\%]) for trauma patients. Patient categories (surgical or medical) do not show significant differences in ventilation time. However, scheduled ICU admissions or surgeries are important factors. Patients with a scheduled ICU admission have a 24.95\% longer ventilation time (CI: [21.89\%, 27.82\%]), while those with a scheduled surgery have a 13.32\% shorter ventilation time (CI: [9.70\%, 16.81\%]) compared to patients without scheduled admissions or surgeries. The referral physician service is also a significant risk factor. Compared with medical referrals, surgical referrals result in a 5.82\% shorter ventilation time (CI: [8.97\%, 2.57\%]), while respirology referrals result in a 16.77\% longer ventilation time (CI: [11.29\%, 22.51\%]).

Specific treatment interventions upon ICU arrival significantly impact ventilation duration. Patients receiving a CVL, AL, IPM, EMO, or IABP have increased ventilation times by 20.20\% (CI: [17.82\%, 22.51\%]), 21.90\% (CI: [19.36\%, 24.48\%]), 70.40\% (CI: [60.16\%, 81.48\%]), 148.93\% (CI: [127.50\%, 172.10\%]), and 40.92\% (CI: [32.45\%, 49.78\%]), respectively. Age is also a significant factor. Compared to patients aged 18 to 39, those aged 40 to 80 and those over 80 have longer ventilation periods, increasing by 14.22\% (CI: [11.29\%, 17.23\%]) and 7.79\% (CI: [4.50\%, 11.18\%]), respectively. 

Both pre-ICU LOS and patients' severity scores, as measured by MODS, are significant risk factors for longer ventilation duration. Longer pre-ICU LOS correlates with longer ventilation times. Patients with a pre-ICU LOS of 2-7 days or $\geq 7$ days experience increases in ventilation duration by 3.87\% (CI: [1.71\%, 6.08\%]) and 12.98\% (CI: [10.52\%, 15.60\%]), respectively, compared to those with $\leq 1$ day pre-ICU LOS. Compared to patients with a MODS score $\leq 1$, those with minimal (1-4), mild (4-8), moderate (8-12), or severe ($>12$) MODS scores experience increases in ventilation duration by 10.63\% (CI: [7.90\%, 13.54\%]), 23.49\% (CI: [20.56\%, 26.49\%]), 30.34\% (CI: [26.49\%, 34.31\%]), and 14.22\% (CI: [7.47\%, 21.53\%]), respectively.

We now turn to the analysis of variance in terms of the variations from individual patients and from ICU sites. Using the mean of the posterior distribution for $\sigma^2_{\gamma}$, the estimated variance ($\sigma^2_{\gamma}$) for the shared ICU site effect from \textit{survregVBfrailty} is 0.1, with a square root of 0.316, which indicates that the average spread of the ventilation time among ICU sites is $\exp(0.316)$, or 1.372. The estimated scale parameter $b$ is 0.444, resulting in an estimated variance of the logarithms of individual patient ventilation times of $0.444^2 \times \pi^2/3$, or 0.649, where $\pi^2/3$ is the variance of a standard log-logistic distribution. To measure the strength of the correlation between patients within the same ICU site, the intra-class correlation coefficient between the logarithms of ventilation time can be estimated by $0.1 / (0.1 + 0.649) \approx 0.134$.

\begin{figure}[!ht]
\begin{subfigure}{.5\textwidth}
  \centering
  \includegraphics[width = 8cm]{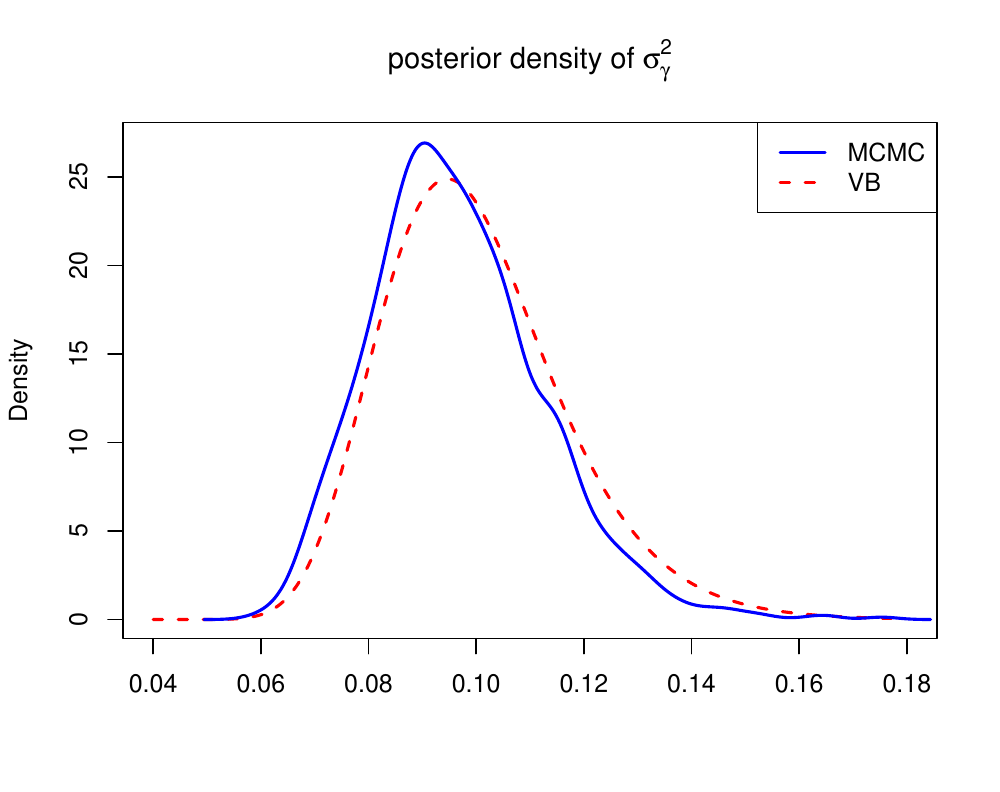}
\end{subfigure} 
\begin{subfigure}{.5\textwidth}
  \centering
  \includegraphics[width = 7.5cm]{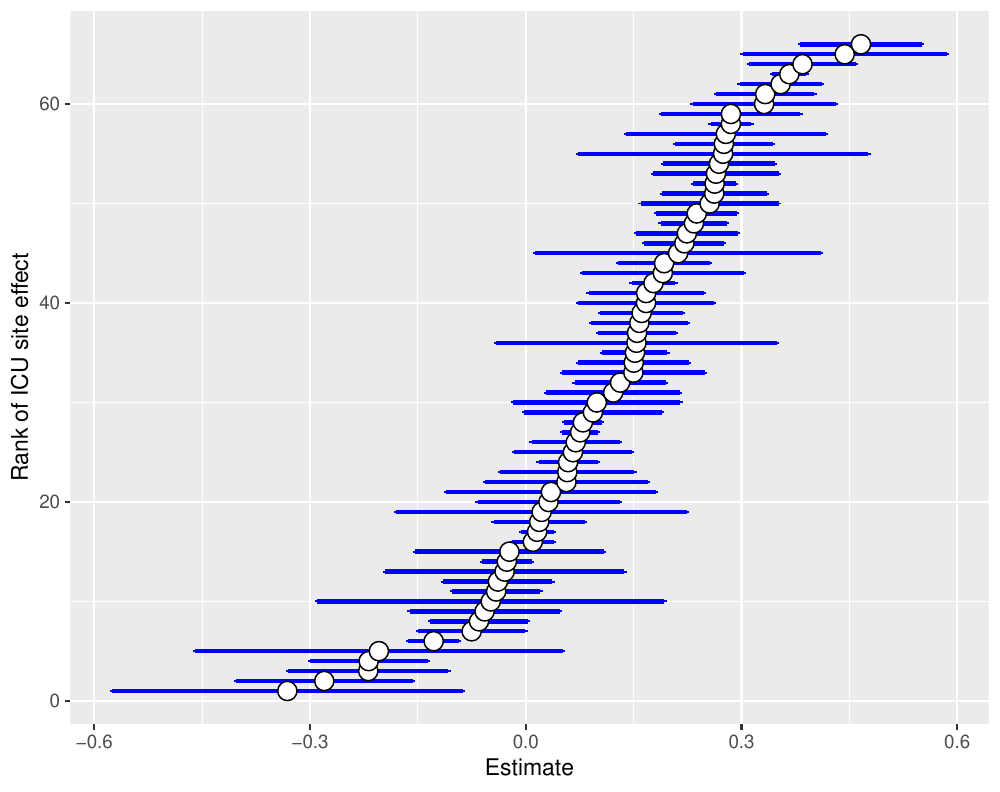}
\end{subfigure}%
\caption{\hspace{0.1cm}\textbf{Left}: Posterior distribution of the variance of the random intercept from VB and MCMC. \textbf{Right}: Estimated ICU site specific random effects with their 95\% credible interval from the proposed VB algorithm. The random effects have been ranked in an increasing order.}
\label{real:random:effect}
\end{figure}

In the left of Figure \ref{real:random:effect}, we observe a strong consistency in estimating the variance of the ICU-site specific random effect between \textit{survregVBfrailty} and the MCMC-based \textit{survregbayes}. We further visualize the estimated ICU-site random effects with the 95\% credible intervals based on their posterior distributions obtained from \textit{survregVBfrailty}, as shown in the right of Figure \ref{real:random:effect}, where the random effects are ranked from smallest to largest. Wider credible intervals correspond to ICUs with larger numbers of patients. As discussed in Lambert et al. \citep{Lambert_2004}, if the intervals overlap, there are no significant center random effects. We observe that some of the intervals do not overlap, indicating that these ICU sites perform differently in terms of patient ventilation duration.

For reference on computational efficiency, the run times in minutes for each method are as follows: \textit{survregVBfrailty} took 1.45 minutes, h-likelihood took 106.18 minutes, MCMC-based \textit{survregbayes} took 267.04 minutes, \textit{survreg} took 0.02 minutes, and \textit{survregVB} took 0.13 minutes.   

\section{Discussion} \label{Sec:Dis}
In this paper, we have proposed a fast variational Bayesian algorithm, called \textit{survregVBfrailty}, for statistical inference using a shared frailty log-logistic AFT model, which can be applied to analyze clustered survival data. We demonstrated that our proposed \textit{survregVBfrailty} algorithm achieves satisfactory estimation performance through simulation studies under various scenarios with different numbers of clusters and cluster sizes. A strong consistency in the estimated posterior distributions was observed between VB and MCMC methods in both the simulation study and the application to ICU ventilation data. Moreover, the \textit{survregVBfrailty} algorithm is significantly more computationally efficient, running over 150 times faster than the MCMC-based \textit{survregbayes} algorithm.

The h-likelihood method proposed by Do Ha et al. \citep{Do_2002} for analyzing clustered survival data in a log-normal AFT model has been shown to be robust against violations of the normality assumption (e.g., extreme value distribution) for the logarithm of survival time. In our simulation study, we investigated the performance of the h-likelihood method when the survival data, given the covariates in a specific cluster, follow a log-logistic distribution. We compared its estimation results to those obtained using VB and MCMC algorithms. We found that the h-likelihood method, \textit{survregHL}, results in higher mean squared errors (MSEs) compared to VB and MCMC algorithms. Specifically, \textit{survregHL} produced a 49.1\% higher MSE than VB and a 41.2\% higher MSE than MCMC for estimating the variance of the random intercept in our simulation study. Additionally, in our application to ICU ventilation duration analysis, we observed significant differences in some regression coefficients between h-likelihood and VB or MCMC methods. Therefore, our proposed shared frailty log-logistic AFT model using variational Bayes can be viewed as a better approach to the shared frailty log-normal AFT model using h-likelihood.

Our application of the proposed method to the CCIS ICU data for ventilation duration analysis reveals a moderate correlation of 0.134 among patients within the same ICU site. By incorporating the ICU site-specific random effect as an unknown shared frailty, we further validate the significant risk factors including the patient severity score MODS identified in the study by Kobara et al. \cite{Kobara_2023} based on the 95\% credible intervals obtained from the posterior distributions of each regression coefficient. We observe a similar trend regarding the effect of MODS on ventilation duration as reported by Kobara et al. \cite{Kobara_2023}. Specifically, when MODS reaches a severe level, the increase in ventilation duration becomes less pronounced. This may be due to patient mortality, which warrants further investigation in future studies. We demonstrate that different ICU sites have varying effects on patient ventilation duration, as observed through differences in the estimated ICU site-specific random effects. Our research provides valuable insights and practical implications for clinical practice and resource management. Specifically, ICUs with smaller estimated random intercepts tend to have shorter overall ventilation durations compared to other ICUs. Understanding the patient characteristics and ventilation practices within these ICUs may help improve clinical performance. For ICUs with larger estimated random intercepts, equipping more ventilators could enhance the overall efficiency of the ICU ventilation procedures.


We present several open problems and directions for future work related to our proposed methodology. As discussed, we assume that the survival time from a specific cluster follows a log-logistic distribution. The robustness of the proposed shared frailty log-logistic AFT model to other distributions remains unknown and warrants further study and comparison with the shared frailty log-normal AFT model. Additionally, extending the current VB algorithm to accommodate other survival distributions, such as the Weibull distribution, could be an interesting avenue of research. In our current framework, we account for cluster-level uncertainty using a random intercept. This approach could be extended to a more general model that includes cluster-level covariates (e.g., the ICU type, general or specialized), resulting in a mixed-effects model. Such a general model can be used to better assess differences in performance across ICU sites regarding the duration of invasive mechanical ventilation. Another potential area for development is the integration of variable selection techniques within the VB algorithm, see Park and Do Ha \cite{Park_2019} as a reference.

\section*{Acknowledgements}
This research is supported by the Natural Sciences and Engineering Research Council of Canada (NSERC). We extend our gratitude to Dr. Yawo M. Kobara for facilitating access to the CCIS data set used in our real data analysis.

\section*{DATA AVAILABILITY STATEMENT}
The ICU data analyzed in Section \ref{sec:real} were from Critical Care Information System (CCIS) Ontario database. Under the authors' data use agreement, they are unable to share the files directly.
\section*{ORCID}
Chengqian Xian: \url{https://orcid.org/0000-0002-6855-3221}

\noindent Camila P.E. de Souza: \url{https://orcid.org/0000-0002-4465-5792}

\noindent Wenqing He: \url{https://orcid.org/0000-0002-8913-9273}

\noindent Felipe F. Rodrigues: \url{https://orcid.org/0000-0001-9370-5998}

\noindent Renfang Tian: \url{https://orcid.org/0000-0001-8176-2469}


\bibliography{wileyNJD-AMA}

\begin{thebibliography}{10}
\providecommand \doibase [0]{http://dx.doi.org/}%

\bibitem{Luo_2013}
Luo S, Yi M, Huang X, Hunt KK. A Bayesian model for misclassified binary outcomes and correlated survival data with applications to breast cancer. {\it Statistics in medicine} 2013\string; 32(13)\string: 2320--2334.

\bibitem{Honerkamp_2016}
Honerkamp-Smith G, Xu R. Three measures of explained variation for correlated survival data under the proportional hazards mixed-effects model. {\it Statistics in medicine} 2016\string; 35(23)\string: 4153--4165.

\bibitem{Liu_207}
Liu XR, Pawitan Y, Clements MS. Generalized survival models for correlated time-to-event data. {\it Statistics in medicine} 2017\string; 36(29)\string: 4743--4762.

\bibitem{Hougaard_1995}
Hougaard P. Frailty models for survival data. {\it Lifetime data analysis} 1995\string; 1\string: 255--273.

\bibitem{Hanagal_2011}
Hanagal DD. {\it Modeling survival data using frailty models}.
\newblock Springer .
\newblock 2011.

\bibitem{Gorfine_2023}
Gorfine M, Zucker DM. Shared frailty methods for complex survival data: a review of recent advances. {\it Annual Review of Statistics and Its Application} 2023\string; 10\string: 51--73.

\bibitem{ICU_2020}
{Canadian Institute for Health Information} . Care in Canadian ICUs. tech. rep., {Canadian Institute for Health Information}; Canada:   2020.

\bibitem{Marshall_1995}
Marshall JC, Cook DJ, Christou NV, Bernard GR, Sprung CL, Sibbald WJ. Multiple organ dysfunction score: a reliable descriptor of a complex clinical outcome. {\it Critical care medicine} 1995\string; 23(10)\string: 1638--1652.

\bibitem{Kobara_2023}
Kobara YM, Wismer M, Rodrigues FF, {de Souza} CPE. Invasive mechanical ventilation duration prediction using survival analysis. {\it International Journal of Healthcare Management} 2023\string: 1--11.

\bibitem{Burgess_2000}
Burgess~Jr JF, Christiansen CL, Michalak SE, Morris CN. Medical profiling: improving standards and risk adjustments using hierarchical models. {\it Journal of health economics} 2000\string; 19(3)\string: 291--309.

\bibitem{Glance_2003}
Glance LG, Dick AW, Osler TM, Mukamel D. Using hierarchical modeling to measure ICU quality. {\it Intensive Care Medicine} 2003\string; 29(12)\string: 2223--2229.
\newblock \href {\doibase 10.1007/s00134-003-1959-9} {doi: 10.1007/s00134-003-1959-9}

\bibitem{Lambert_2004}
Lambert P, Collett D, Kimber A, Johnson R. Parametric accelerated failure time models with random effects and an application to kidney transplant survival. {\it Statistics in medicine} 2004\string; 23(20)\string: 3177--3192.

\bibitem{Do_2002}
Do~Ha I, Lee Y, Song JK. Hierarchical-likelihood approach for mixed linear models with censored data. {\it Lifetime data analysis} 2002\string; 8\string: 163--176.

\bibitem{Do_2017}
Do~Ha I, Jeong JH, Lee Y. Statistical modelling of survival data with random effects. {\it Statistics for Biology and Health} 2017.

\bibitem{Do_2012}
Do~Ha I, Noh M, Lee Y. frailtyHL: A Package for Fitting Frailty Models with H-likelihood.. {\it R J.} 2012\string; 4(2)\string: 28.

\bibitem{Park_2019}
Park E, Do~Ha I. Penalized variable selection for accelerated failure time models with random effects. {\it Statistics in medicine} 2019\string; 38(5)\string: 878--892.

\bibitem{Zhou_2017}
Zhou H, Hanson T, Zhang J. Generalized accelerated failure time spatial frailty model for arbitrarily censored data. {\it Lifetime data analysis} 2017\string; 23\string: 495--515.

\bibitem{Zhou_2020}
Zhou H, Hanson T, Zhang J. spBayesSurv: Fitting Bayesian Spatial Survival Models Using R. {\it Journal of Statistical Software} 2020\string; 92(9)\string: 1–33.
\newblock \href {\doibase 10.18637/jss.v092.i09} {doi: 10.18637/jss.v092.i09}

\bibitem{Bishop_2006}
Bishop C. {\it {Pattern Recognition and Machine Learning}}.
\newblock Springer .
\newblock 2006.

\bibitem{Hoffman_2013}
Hoffman MD, Blei DM, Wang C, Paisley J. Stochastic variational inference. {\it Journal of Machine Learning Research} 2013\string; 14\string: 1303-1347.

\bibitem{Ranganath_2014}
Ranganath R, Gerrish S, Blei D. Black box variational inference. In: PMLR. ; 2014\string: 814--822.

\bibitem{Jordan_1999}
Jordan MI, Ghahramani Z, Jaakkola T, Saul L. {Introduction to variational methods for graphical models}. {\it Machine Learning} 1999\string; 37\string: 183–233.

\bibitem{Neville_2011}
Neville SE, Palmer M, Wand M. Generalized extreme value additive model analysis via mean field variational bayes. {\it Australian \& New Zealand Journal of Statistics} 2011\string; 53(3)\string: 305--330.

\bibitem{Pham_2013}
Pham TH, Ormerod JT, Wand M. {Mean field variational Bayesian inference for nonparametric regression with measurement error}. {\it Computational Statistics \& Data Analysis} 2013\string; 68\string: 375-387.
\newblock \href {\doibase https://doi.org/10.1016/j.csda.2013.07.014} {doi: https://doi.org/10.1016/j.csda.2013.07.014}

\bibitem{Luts_2015}
Luts J, Wand MP. {Variational inference for count response semiparametric regression}. {\it Bayesian Analysis} 2015\string; 10(4)\string: 991 -- 1023.
\newblock \href {\doibase 10.1214/14-BA932} {doi: 10.1214/14-BA932}

\bibitem{Ray_2022}
Ray K, Szabó B. {Variational Bayes for high-dimensional linear regression with sparse priors}. {\it Journal of the American Statistical Association} 2022\string; 117(539)\string: 1270-1281.
\newblock \href {\doibase 10.1080/01621459.2020.1847121} {doi: 10.1080/01621459.2020.1847121}

\bibitem{Lee_2016}
Lee CYY, Wand MP. Streamlined mean field variational Bayes for longitudinal and multilevel data analysis. {\it Biometrical Journal} 2016\string; 58(4)\string: 868--895.

\bibitem{Xian_clu_2024}
Xian C, {de Souza} CPE, Jewell J, Dias R. Clustering functional data via variational inference. {\it Advances in Data Analysis and Classification} 2024\string: 1--50.

\bibitem{daCruz_2024}
{da Cruz} AC, {de Souza} CPE, Sousa PHTO. Fast Bayesian Basis Selection for Functional Data Representation with Correlated Errors. {\it arXiv} 2024.

\bibitem{David_2017}
Blei DM, Kucukelbir A, McAuliffe JD. {Variational inference: A review for statisticians}. {\it Journal of the American Statistical Association} 2017\string; 112(518)\string: 859-877.

\bibitem{Xian_2024}
Xian C, {de Souza} CPE, He W, Rodrigues FF, Tian R. Variational Bayesian analysis of survival data using a log-logistic accelerated failure time model. {\it Statistics and Computing} 2024\string; 34(2)\string: 67.

\bibitem{Robinson_1991}
Robinson GK. That BLUP is a good thing: the estimation of random effects. {\it Statistical science} 1991\string: 15--32.

\bibitem{Nolan_2020}
Nolan TH, Menictas M, Wand MP. Streamlined variational inference with higher level random effects. {\it Journal of Machine Learning Research} 2020\string; 21(157)\string: 1--62.

\bibitem{LeeCathy_2016}
Lee CYY, Wand MP. Variational methods for fitting complex Bayesian mixed effects models to health data. {\it Statistics in medicine} 2016\string; 35(2)\string: 165--188.

\bibitem{Yao_2018}
{\it Yes, but did it Work?: Evaluating variational inference}. 80 of {\it Proceedings of Machine Learning Research}; PMLR:   2018.

\end{thebibliography}
\section*{Appendix A. Update equations}\label{Append.A}
\noindent\textbf{\textit{(1) Update equation for $q^*(\negr{\beta})$}}
\begin{eqnarray}
\log q^*(\negr{\beta}) \addeq \E_{-\negr{\beta}}[\log p(\vect{D}\,\vert\,\negr{\beta}, \negr{\gamma}, b)]+\E_{-\negr{\beta}}[\log p(\negr{\beta})] \nonumber
\end{eqnarray}
where
\begin{eqnarray}
&& \E_{-\negr{\beta}}[\log p(\vect{D}\,\vert\,\negr{\beta}, \negr{\gamma}, b)] \nonumber \\
&=& \E_{-\negr{\beta}}\Bigg[-\delta\log b +\sum_{i=1}^K\sum_{j=1}^{n_i} \Bigg\{\delta_{ij} \frac{y_{ij}-\vect{X}_{ij}^T\negr{\beta}-\gamma_i}{b} -(1+\delta_{ij})\log\Big\{1+\exp\big(\frac{y_{ij}-\vect{X}_{ij}^T \negr{\beta}-\gamma_i}{b}\big)\Big\}\Bigg\}\Bigg] \label{eq:beta:1}
\end{eqnarray}
with $\delta=\sum_{i=1}^K\sum_{j=1}^{n_i}\delta_{ij}$ being the number of observed survival times, and
\begin{eqnarray}
\E_{-\negr{\beta}}[\log p(\negr{\beta})]  &\addeq& \frac{p}{2}\log v_0-\frac{1}{2}v_0(\negr{\beta}-\negr{\mu}_0)^T(\negr{\beta}-\negr{\mu}_0)\nonumber\\  &\addeq& -\frac{1}{2}v_0\,\big[\negr{\beta}^T\negr{\beta}-2\negr{\mu}_0^T\negr{\beta}\big] = v_0\,\negr{\mu}_0^T\negr{\beta} -\frac{1}{2}v_0\negr{\beta}^T\negr{\beta} \nonumber
\end{eqnarray}
In Equation (\ref{eq:beta:1}), we use the quadratic piece-wise approximation proposed by Xian et al to approximate $1+\exp\big(\frac{y_{ij}-\vect{X}_{ij}^T \negr{\beta}-\gamma_i}{b}\big)$:
\begin{eqnarray}
1+\exp\big(\frac{y_{ij}-\vect{X}_{ij}^T \negr{\beta}-\gamma_i}{b}\big)
\approx \rho_{ij}\frac{y_{ij}-\vect{X}_{ij}^T \negr{\beta}-\gamma_i}{b} + \zeta_{ij}\big(\frac{y_{ij}-\vect{X}_{ij}^T \negr{\beta}-\gamma_i}{b}\big)^2 \label{eq.quadratic.approx}
\end{eqnarray}
where $\rho_{ij}:=0^{\nu_{ij1}}\times 0.1696^{\nu_{ij2}}\times 0.5^{\nu_{ij3}}\times 0.8303^{\nu_{ij4}}\times 1^{{1-\sum_{k=1}^4\nu_{ijk}}}$ and $\zeta_{ij}:=0^{\nu_{ij1}}\times 0.0189^{\nu_{ij2}}\times 0.1138^{\nu_{ij3}}\times 0.0190^{\nu_{ij4}}\times 0^{{1-\sum_{k=1}^4\nu_{ijk}}}$ with
$\nu_{ij1}=\mathbbm{1}\big(\frac{y_{ij}-\vect{X}_{ij}^T\negr{\beta}-\gamma_i}{b} \leq -5\big)$, $\nu_{ij2}=\mathbbm{1}\big(-5 < \frac{y_{ij}-\vect{X}_{ij}^T\negr{\beta}-\gamma_i}{b} \leq -1.7\big)$, $\nu_{ij3}=\mathbbm{1}\big(-1.7 <\frac{y_{ij}-\vect{X}_{ij}^T\negr{\beta}-\gamma_i}{b} \leq 1.7\big)$, and 
$\nu_{ij4}=\mathbbm{1}\big(1.7 <\frac{y_{ij}-\vect{X}_{ij}^T\negr{\beta}-\gamma_i}{b} \leq 5\big)$.
Then, we obtain,
\begin{eqnarray}
\log q^*(\negr{\beta}) &\addeq& \Bigg[v_0\,\negr{\mu}_0^T + \bigg(\sum_{i=1}^K\sum_{j=1}^{n_i} \Big(\E_{q(b)}\big(\frac{1}{b}\big)\Big(-\delta_{ij}+(1+\delta_{ij})\rho_{ij}\Big) +2\E_{q(b)}\big(\frac{1}{b^2}\big) (1+\delta_{ij})\zeta_{ij}\big(y_{ij}-\E_{q(\gamma_i)}(\gamma_i)\big) \Big)\vect{X}_{ij}^T\bigg)\Bigg]\,\negr{\beta} \nonumber\\
&& \,-\frac{1}{2}\negr{\beta}^T\bigg(v_0\textbf{I}+2\E_{q(b)}\big(\frac{1}{b^2}\big) \sum_{i=1}^K\sum_{j=1}^{n_i}(1+\delta_{ij})\zeta_{ij}\vect{X}_{ij}\vect{X}_{ij}^T\bigg)\,\negr{\beta}\nonumber
\end{eqnarray}
Let 
\begin{eqnarray}
\Sigma^*:= \bigg[v_0\textbf{I}+2\E_{q(b)}\big(\frac{1}{b^2}\big) \sum_{i=1}^K\sum_{j=1}^{n_i}(1+\delta_{ij})\zeta_{ij}\vect{X}_{ij}\vect{X}_{ij}^T\bigg]^{-1} \nonumber
\label{eq_update_sigma}
\end{eqnarray}
and 
\begin{eqnarray}
\negr{\mu}^*:=\Bigg[\bigg\{v_0\,\negr{\mu}_0^T + \sum_{i=1}^K\sum_{j=1}^{n_i} \Big(\E_{q(b)}\big(\frac{1}{b}\big)\Big(-\delta_{ij}+(1+\delta_{ij})\rho_{ij}\Big) +2\E_{q(b)}\big(\frac{1}{b^2}\big) (1+\delta_{ij})\zeta_{ij}\big(y_{ij}-\E_{q(\gamma_i)}(\gamma_i)\big) \Big)\vect{X}_{ij}^T\bigg\}\,\Sigma^{*}\Bigg]^{T} \nonumber
\label{eq_update_mu}
\end{eqnarray}
Then, $q^*(\negr{\beta})$ is $N_p(\negr{\mu}^{*}, \Sigma^*)$. \vspace{0.2cm}

\noindent\textbf{\textit{(2) Update equation for $q^*({\gamma}_i)$}}
\begin{eqnarray}
\log q^*({\gamma}_i) \addeq \E_{-{\gamma}_i}\Big[\log p(\vect{D}\,\vert\,\negr{\beta}, \negr{\gamma}, b)\Big]+\E_{-{\gamma}_i}\Big[\sum_{i=1}^K \log p({\gamma}_i\, \vert \sigma^2_{\gamma})\Big] \nonumber
\end{eqnarray}
where
\begin{eqnarray}
&&\E_{-{\gamma}_i}\Big[\log p(\vect{D}\,\vert\,\negr{\beta}, \negr{\gamma}, b)\Big]  \nonumber\\
&\addeq& \E_{-{\gamma}_i}\Bigg[\sum_{j=1}^{n_i} \Big\{\delta_{ij} \frac{y_{ij}-\vect{X}_{ij}^T\negr{\beta}-\gamma_i}{b} -(1+\delta_{ij})\log\Big(1+\exp\big(\frac{y_{ij}-\vect{X}_{ij}^T \negr{\beta}-\gamma_i}{b}\big)\Big)\Big\} \Bigg] \label{eq:gamma.update.1}
\end{eqnarray}
and 
\begin{eqnarray}
\E_{-{\gamma}_i}\Big[\sum_{i=1}^K \log p({\gamma}_i\, \vert \sigma^2_{\gamma})\Big] \addeq -\frac{1}{2}\gamma_i^2\E_{q(\sigma^2_{\gamma})}\big(\frac{1}{\sigma^2_{\gamma}}\big)  \nonumber
\end{eqnarray}
We again apply the quadratic approximation in (\ref{eq.quadratic.approx}) to (\ref{eq:gamma.update.1}), and obtain 
\begin{eqnarray}
\log q^*({\gamma}_i) &\addeq& \gamma_i \Bigg(\sum_{j=1}^{n_i}\Big[\E_{q(b)}(\frac{1}{b})\big(-\delta_{ij}+(1+\delta_{ij})\rho_{ij}\big) +2\E_{q(b)}(\frac{1}{b^2})(1+\delta_{ij})\zeta_{ij}\big(y_{ij}-\vect{X}_{ij}^T\E_{q(\negr{\beta})}\negr{\beta}\big)\Big]\Bigg)
\nonumber\\
&& \,-\frac{1}{2} \gamma_i^2\bigg(\E_{q(\sigma^2_{\gamma})}\big(\frac{1}{\sigma^2_{\gamma}}\big)+2\E_{q(b)}\big(\frac{1}{b^2}\big)\sum_{j=1}^{n_i}(1+\delta_{ij})\zeta_{ij}\bigg) \nonumber
\end{eqnarray}
Let 
\begin{eqnarray}
\sigma^{2*}_{i}= \big[\E_{q(\sigma^2_{\gamma})}\big(\frac{1}{\sigma^2_{\gamma}}\big)+2\E_{q(b)}\big(\frac{1}{b^2}\big)\sum_{j=1}^{n_i}(1+\delta_{ij})\zeta_{ij}\big]^{-1}\nonumber
\end{eqnarray}
and 
\begin{eqnarray}
\tau^*_{i}=\sigma^{2*}_{i} \sum_{j=1}^{n_i}\Big[\E_{q(b)}(\frac{1}{b})\big(-\delta_{ij}+(1+\delta_{ij})\rho_{ij}\big) +2\E_{q(b)}(\frac{1}{b^2})(1+\delta_{ij})\zeta_{ij}\big(y_{ij}-\vect{X}_{ij}^T\E_{q(\negr{\beta})}\negr{\beta}\big)\Big]  \nonumber
\end{eqnarray}
Then, $q^*({\gamma}_i)$ is $N_l(\tau_i^*, \sigma_i^{2*})$. \vspace{0.2cm}

\noindent\textbf{\textit{(3) Update equation for $q^*(b)$}}
\begin{eqnarray}
\log q^*(b) \addeq \E_{-b}[\log p(\vect{D}\,\vert\,\negr{\beta}, \negr{\gamma}, b)]+\E_{-b}[\log p(b)] \nonumber
\end{eqnarray}
where
\begin{eqnarray}
&&\E_{-b}[\log p(\vect{D}\,\vert\,\negr{\beta}, \negr{\gamma}, b)]\nonumber\\ 
&=& \E_{-b}\Bigg[-\delta\log b +\sum_{i=1}^K\sum_{j=1}^{n_i} \Bigg\{\delta_{ij} \frac{y_{ij}-\vect{X}_{ij}^T\negr{\beta}-\gamma_i}{b} -(1+\delta_{ij})\log\Big\{1+\exp\big(\frac{y_{ij}-\vect{X}_{ij}^T \negr{\beta}-\gamma_i}{b}\big)\Big\}\Bigg\}\Bigg] \label{eq:update.b.1}
\end{eqnarray}
and
\begin{eqnarray}
\E_{-b}[\log p(b)] \addeq -(\alpha_0+1)\log b - \frac{\omega_0}{b}\quad  \text{since we assume the prior}\;  b \sim \text{Inverse-Gamma}\,(\alpha_0, \omega_0)\nonumber 
\end{eqnarray}
In Equation (\ref{eq:update.b.1}), we apply the linear piece-wise approximation proposed by Xian et al. to approximate $1+\exp\big(\frac{y_{ij}-\vect{X}_{ij}^T \negr{\beta}-\gamma_i}{b}\big)$ to achieve conjugacy:
\begin{eqnarray}
1+\exp\big(\frac{y_{ij}-\vect{X}_{ij}^T \negr{\beta}-\gamma_i}{b}\big) \approx \varphi_{ij} \frac{y_{ij}-\vect{X}_{ij}^T \negr{\beta}-\gamma_i}{b}. \nonumber
\end{eqnarray}
where $\varphi_{ij} = 0^{\eta_{ij1}}\times 0.0426^{\eta_{ij2}}\times 0.3052^{\eta_{ij3}}\times 0.6950^{\eta_{ij4}} \times\, 0.9574^{\eta_{ij5}}
 \times  1^{1-\sum_{k=1}^5\eta_{ijk}}$ with $\eta_{ij1} = \mathbbm{1}\big(\frac{y_{ij}-\vect{X}_{ij}^T\negr{\beta}-\gamma_i}{b} \leq -5\big)$, $\eta_{ij2} = \mathbbm{1}\big(-5 < \frac{y_{ij}-\vect{X}_{ij}^T\negr{\beta}-\gamma_i}{b} \leq -1.701\big)$, $\eta_{ij3} = \mathbbm{1}\big(-1.701 < \frac{y_{ij}-\vect{X}_{ij}^T\negr{\beta}-\gamma_i}{b} \leq 0\big)$, $\eta_{ij4} = \mathbbm{1}\big(0 < \frac{y_{ij}-\vect{X}_{ij}^T\negr{\beta}-\gamma_i}{b} \leq 1.702\big)$, and $\eta_{ij5} = \mathbbm{1}\big(1.702 < \frac{y_{ij}-\vect{X}_{ij}^T\negr{\beta}-\gamma_i}{b} \leq 5\big)$.
 
Then, we obtain
\begin{eqnarray}
\log q^*(b) \addeq -(\alpha_0+\delta + 1)\log b - \frac{1}{b}\Big[\omega_0-\sum_{i=1}^K\sum_{j=1}^{n_i}\Big(\delta_{ij}-(1+\delta_{ij})\varphi_{ij}\Big)\Big(y_{ij}-\vect{X}_{ij}^T\E_{q(\negr{\beta})}(\negr{\beta})-\E_{q(\gamma_i)}\gamma_i\Big)\Big]\nonumber
\end{eqnarray}

Let $\alpha^* = \alpha_0 + \delta$ and
\begin{eqnarray}
\omega^* = \omega_0-\sum_{i=1}^K\sum_{j=1}^{n_i}\Big(\delta_{ij}-(1+\delta_{ij})\varphi_{ij}\Big)\Big(y_{ij}-\vect{X}_{ij}^T\E_{q(\negr{\beta})}(\negr{\beta})-\E_{q(\gamma_i)}\gamma_i\Big),\nonumber
\end{eqnarray}
we have $q^*(b)$ is $\text{Inverse-Gamma}\,(\alpha^*, \omega^*)$. \vspace{0.2cm}

\noindent\textbf{\textit{(4) Update equation for $q^*(\sigma^2_{\gamma})$}}
\begin{eqnarray}
\log q^*(\sigma^2_{\gamma}) \addeq \E_{-\sigma^2_{\gamma}}\Big[\sum_{i=1}^K \log p({\gamma}_i\, \vert \sigma^2_{\gamma})\Big]+\E_{-\sigma^2_{\gamma}}\Big[\log p(\sigma^2_{\gamma})\Big]\nonumber
\end{eqnarray}
where
\begin{eqnarray}
\E_{-\sigma^2_{\gamma}}\Big[\sum_{i=1}^K \log p({\gamma}_i\, \vert \sigma^2_{\gamma})\Big] &=&  \E_{-\sigma^2_{\gamma}}\bigg[\sum_{i=1}^K \log\Big(\frac{1}{\sqrt{2\pi \sigma^2_{\gamma}}}\exp(-\frac{1}{2\sigma^2_{\gamma}}\gamma_i^2)\Big) \bigg]\nonumber\\
&\addeq& -\frac{K}{2}\log \sigma^2_{\gamma} -\frac{1}{2\sigma^2_{\gamma}}\sum_{i=1}^K\E_{q(\gamma_i)}\gamma_i^2 \nonumber 
\end{eqnarray}
and $\E_{-\sigma^2_{\gamma}}\big[\log p(\sigma^2_{\gamma})\big] = -(\lambda_0 + 1)\log \sigma^2_{\gamma} - \eta_0/{\sigma^2_{\gamma}}$.

Therefore,
\begin{eqnarray}
\log q^*(\sigma^2_{\gamma}) \addeq -(\lambda_0 + \frac{K}{2} + 1)\log \sigma^2_{\gamma} - \big[\eta_0 + \frac{1}{2}\sum_{i=1}^K\E_{q(\gamma_i)}\gamma_i^2\big] / {\sigma^2_{\gamma}} \nonumber
\end{eqnarray}

Let $\lambda^*=\lambda_0 + K/2$ and
\begin{eqnarray}
\eta^* = \eta_0 + \frac{1}{2}\sum_{i=1}^K\E_{q(\gamma_i)}\gamma_i^2, \nonumber
\end{eqnarray}
$q^*(\sigma^2_{\gamma})$ is an $\text{Inverse-Gamma}\,(\lambda^*, \eta^*)$.
\section*{Appendix B. ELBO calculation}\label{Append.B}
We now present details of calculating $\textit{diff}_{\negr{\beta}}+\textit{diff}_{\negr{\gamma}}+\textit{diff}_{b}+\textit{diff}_{\sigma^2_{\gamma}}$ in the ELBO defined in (\ref{VBAFT.frailty.eq:ELBO}).
\begin{eqnarray}
\textit{diff}_{\negr{\beta}} &=&\E_{q}[\log p(\negr{\beta})] - \E_{q}[\log q(\negr{\beta})] \nonumber \\
&\addeq& \E_{q}[-\frac{1}{2}v_0(\negr{\beta}-\negr{\mu}_0)^T(\negr{\beta}-\negr{\mu}_0)] -\E_{q}[-\frac{1}{2}\log (\vert\Sigma^*\vert)-\frac{1}{2}(\negr{\beta}-\negr{\mu}^*)^T(\Sigma^*)^{-1}(\negr{\beta}-\negr{\mu}^*)] \nonumber \\
&\addeq& -\frac{1}{2}v_0[\text{trace}(\Sigma^*)+(\negr{\mu}^*-\negr{\mu}_0)^T(\negr{\mu}^*-\negr{\mu}_0)] + \frac{1}{2}\log (\vert\Sigma^*\vert). \nonumber
\end{eqnarray}
Note that
\begin{eqnarray}
\E_{q}[\frac{1}{2}(\negr{\beta}-\negr{\mu}^*)^T(\Sigma^*)^{-1}(\negr{\beta}-\negr{\mu}^*)]=\frac{p}{2}, \nonumber
\end{eqnarray}
which is always a constant at each iteration and therefore we ignore it. 
\begin{eqnarray}
\textit{diff}_{\negr{\gamma}} &=&\E_{q}\big[\sum_{i=1}^K \log p(\gamma_i\, \vert \sigma_{\gamma}^2)\big]-\E_{q}\big[\sum_{i=1}^K \log q(\gamma_i)\big]\nonumber\\
&\addeq& -\frac{K}{2}\E_{q(\sigma^2_\gamma)}(\log \sigma^2_\gamma)-\frac{1}{2}\E_{q(\sigma^2_\gamma)}\big(\frac{1}{\sigma^2_\gamma}\big)\sum_{i=1}^K \E_{q(\gamma_i)}(\gamma_i^2)  -\frac{1}{2} \sum_{i=1}^K\big[\log \sigma_i^{2*}-\frac{1}{\sigma_i^{2*}}\big(\E_{q(\gamma_i)}(\gamma_i^2)-\mu_i^{*2}\big)\big]\nonumber\\
&\addeq& -\frac{K}{2}\E_{q(\sigma^2_\gamma)}(\log \sigma^2_\gamma)-\frac{1}{2}\E_{q(\sigma^2_\gamma)}\big(\frac{1}{\sigma^2_\gamma}\big)\sum_{i=1}^K \E_{q(\gamma_i)}(\gamma_i^2) -\frac{1}{2} \sum_{i=1}^K\big(\log \sigma_i^{2*}\big)\nonumber
\end{eqnarray}
Since $\big(\E_{q(\gamma_i)}(\gamma_i^2)-\mu_i^{*2}\big)/\sigma_i^{2*}=1$ for $i = 1,... K$, and therefore,
$$\Big(\sum_{i=1}^K \big(\E_{q(\gamma_i)}(\gamma_i^2)-\mu_i^{*2}\big)\Big)/ (2\sigma_i^{2*})=K/2$$
which is a constant, and we can ignore it.

\begin{eqnarray}
\textit{diff}_{b}&=&\E_{q}[\log p(b)]-\E_{q}[\log q(b)]\nonumber \\
&\addeq& \E_{q}\Big[-(\alpha_0+1)\log b-\frac{\omega_0}{b}\Big] \nonumber\\
&&\,- \E_{q}\Big[\alpha^*\log \omega^* - \log (\Gamma(\alpha^*))-(\alpha^*+1)\log b-\frac{\omega^*}{b}\Big] \nonumber \\
&=& (\alpha^*-\alpha_0)\E_{q(b)}(\log b) + (\omega^*-\omega_0)\E_{q(b)}\big(\frac{1}{b}\big) - \alpha^*\log \omega^*. \nonumber
\end{eqnarray}
Since $\alpha^*$ does not change at each iteration, we remove $\log (\Gamma(\alpha^*))$ in the calculation of the ELBO.
\begin{eqnarray}
\textit{diff}_{\sigma^2_{\gamma}}&=&\E_{q}[\log p(\sigma^2_{\gamma})]-\E_{q}[\log q(\sigma^2_{\gamma})] \nonumber \\
&\addeq& \E_{q}\Big[-(\lambda_0+1)\log \sigma^2_{\gamma}-\frac{\eta_0}{\sigma^2_{\gamma}}\Big] \nonumber\\
&&\,- \E_{q}\Big[\lambda^*\log \eta^* - \log (\Gamma(\lambda^*))-(\lambda^*+1)\log \sigma^2_{\gamma}-\frac{\eta^*}{\sigma^2_{\gamma}}\Big] \nonumber \\
&=& (\lambda^*-\lambda_0)\E_{q(\sigma^2_{\gamma})}(\log \sigma^2_{\gamma}) + (\eta^*-\eta_0)\E_{q(\sigma^2_{\gamma})}\big(\frac{1}{\sigma^2_{\gamma}}\big) - \lambda^*\log \eta^*. \nonumber
\end{eqnarray}
Since $\lambda^*$ does not change at each iteration, we remove $\log (\Gamma(\lambda^*))$ in the calculation of the ELBO.

\section*{Appendix C. Real data analysis}\label{Append.C}
\begin{landscape}
\begin{table}[!ht]
\begin{center}
\caption{\hspace{0.1cm} \textit{Results for ICU ventilation duration analysis:} Posterior means (Mean) with 95\% credible intervals (95\% Cred. Int.) 
from the VB algorithms and MCMC-based \textit{survregbayes}, respectively. Point estimates (Est.) with 95\% confidence interval (95\% Conf. Int.) from the h-likelihood \textit{survregHL} and \textit{survreg} methods.}
\label{tab.results.real}
\begin{tabular}{llllllllllllllllll}
\toprule
\multicolumn{1}{l}{} & \multicolumn{6}{c}{With frailty} &  \multicolumn{4}{c}{Without frailty}\\
\cmidrule(rl){2-7} \cmidrule(rl){8-11} 
\multicolumn{1}{l}{} & \multicolumn{2}{c}{\textit{survregVBfrailty}} &  \multicolumn{2}{c}{\textit{survregHL}} & \multicolumn{2}{c}{\textit{survregbayes}}& \multicolumn{2}{c}{\textit{survreg}}& \multicolumn{2}{c}{\textit{survregVB}} \\
\cmidrule(rl){2-3} \cmidrule(rl){4-5} \cmidrule(rl){6-7}\cmidrule(rl){8-9}\cmidrule(rl){10-11}
 & Mean  & 95\% Cred. Int. & Est.  & 95\% Conf. Int. & Mean  & 95\% Cred. Int. & Est.  & 95\% Conf. Int. & Mean  & 95\% Cred. Int.\\
\midrule
 Admission Source &  &  &  &  &   &  &  &  &   &\\
 \quad Downstream (ref)& - & - &  -&-  & -  &- & - & - &  - &-\\
 \quad ED & -0.134 & [-0.171, -0.097] & -0.134 & [-0.181, -0.086] & -0.134 & [-0.172, -0.103] & -0.135 & [-0.173, -0.098] & -0.134  & [-0.171, -0.097]\\
 \quad Home & 0.101 & [-0.009, 0.211] & 0.088 & [-0.016, 0.191] & 0.096 & [-0.014, 0.211] & 0.095 & [-0.011, 0.201] & 0.092  & [-0.013, 0.197]\\
 \quad Hospital & 0.098 & [0.068, 0.127] & 0.084 & [0.033, 0.135] & 0.097 & [0.064, 0.128]  & 0.077 & [0.037, 0.117] & 0.079  & [0.039, 0.119]\\
 \quad OR & -0.327 & [-0.357, -0.296] & -0.319 & [-0.372, -0.266] & -0.312 & [-0.342, -0.281] & -0.333 & [-0.374, -0.292] &  -0.334 & [-0.375, -0.293] \\
 \quad Other & 0.122 & [0.050, 0.193] & 0.062 & [-0.062, 0.186] & 0.115 & [0.039, 0.191]  & 0.146 & [0.043, 0.248] & 0.151  & [0.049, 0.252] \\
 \quad Ward & -0.140& [-0.178, -0.102] & -0.142 & [-0.191, -0.094] & -0.140 & [-0.178, -0.101]& -0.135 & [-0.174, -0.097] & -0.137  & [-0.175, -0.099]\\\midrule
 Diagnosis  &  &  &  &  &   & &  &  &   &\\
 \quad Cardiovascular (ref)& - & - &  -&-  & -  &-  &  &  &   &\\
 \quad Gastrointestinal  & 0.260 & [0.226, 0.294] & 0.265 & [0.219, 0.310] & 0.260 & [0.226, 0.294]  & 0.265 & [0.231, 0.300] &  0.265 & [0.230, 0.299]\\
 \quad Neurological & 0.286 & [0.255, 0.318] & 0.286 & [0.244, 0.329] & 0.280 & [0.250, 0.317]  & 0.304 & [0.272, 0.336] & 0.303  & [0.270, 0.335]\\
 \quad Other & 0.379 & [0.356, 0.402] & 0.366 & [0.335, 0.397] & 0.370 & [0.349, 0.394] & 0.386 & [0.363, 0.410] &  0.386 & [0.363, 0.410]\\
 \quad Trauma  & 0.575 & [0.529, 0.621] & 0.527 & [0.467, 0.586] & 0.574 & [0.530, 0.622] & 0.585 & [0.539, 0.631] &  0.589 & [0.543, 0.635]\\\midrule
 Patient type  &  &  &  &  &   &  &  &  &   &\\
 \quad Medical (ref)& - & - &  -&-  & -  &-  &  &  &   &\\
 \quad Surgical  & -0.021 & [-0.060, 0.017] & -0.019 & [-0.031, 0.069] & -0.023 & [-0.058, 0.013] & -0.022 & [-0.061, 0.017] & -0.024  &[-0.064, 0.015]\\
 \midrule
 Scheduled Admission  &  &  &  &  &   &  &  &  &   &\\
 \quad No (ref)& - & - &  -&-  & -  &-  \\
 \quad Yes & -0.287 & [-0.326, -0.247] & -0.281 & [-0.336, -0.226] & -0.285 & [-0.329, -0.246]  & -0.289 & [-0.329, -0.249] & -0.277  &[-0.317, -0.237]\\\midrule
 Scheduled Surgery  &  &  &  &  &   &  &  &  &   &\\
 \quad No (ref)& - & - &  -&-  & -  &-  &  &  &   &\\
 \quad Yes & -0.143 & [-0.184, -0.102] & -0.162 & [-0.218, -0.106] & -0.146 & [-0.188, -0.104]  & -0.141 & [-0.183, -0.099] & -0.144  &[-0.186, -0.102]\\
 \midrule
 Referral  &  &  &  &  &   & &  &  & &   \\
 \quad Medical (ref)& - & - &  -&-  & -  &-  &  &  & & \\
 \quad Other   & -0.028 & [-0.058, 0.001] & -0.012 & [-0.052, 0.027] & -0.034 & [-0.065, -0.004] & -0.051 & [-0.080, -0.021] & -0.052 & [-0.081, -0.022] \\
 \quad Respirology & 0.155 & [0.107, 0.203] & 0.112 & [0.047, 0.177] & 0.156 & [0.108, 0.205] & 0.165  & [0.117, 0.214] & 0.168 & [0.120, 0.216] \\
 \quad Surgical  & -0.060 & [-0.094, -0.026] & -0.052 & [-0.098, -0.005] & -0.061 & [-0.099, -0.024]  & -0.048 & [-0.083, -0.013] & -0.049  & [-0.085, -0.014]\\
\toprule
\end{tabular}
\end{center}
\end{table}
\end{landscape}

\begin{landscape}
\begin{table}[!ht]
\begin{center}
\caption*{\textbf{TABLE 4 Continued} \textit{Results for ICU ventilation duration analysis:} Posterior means (Mean) 
with 95\% credible intervals (95\% Cred. Int.) 
from the VB algorithms and MCMC-based \textit{survregbayes}, respectively. Point estimates (Est.) with 95\% confidence interval (95\% Conf. Int.) from the h-likelihood \textit{survregHL} and \textit{survreg} methods.}
\begin{tabular}{llllllllllllllllll}
\toprule
\multicolumn{1}{l}{} & \multicolumn{6}{c}{With frailty} &  \multicolumn{4}{c}{Without frailty}\\
\cmidrule(rl){2-7} \cmidrule(rl){8-11} 
\multicolumn{1}{l}{} & \multicolumn{2}{c}{\textit{survregVBfrailty}} &  \multicolumn{2}{c}{\textit{survregHL}} & \multicolumn{2}{c}{\textit{survregbayes}}& \multicolumn{2}{c}{\textit{survreg}}& \multicolumn{2}{c}{\textit{survregVB}} \\
\cmidrule(rl){2-3} \cmidrule(rl){4-5} \cmidrule(rl){6-7}\cmidrule(rl){8-9}\cmidrule(rl){10-11}
 & Mean  & 95\% Cred. Int. & Est.  & 95\% Conf. Int. & Mean  & 95\% Cred. Int. & Est.  & 95\% Conf. Int. & Mean  & 95\% Cred. Int.\\
\midrule
 Other Interventions  &  &  &  &  &   & &  &  & &  \\
 \quad No (ref)& - & - &  -&-  & -  &-  &  &  & & \\
 \quad Yes  & 0.088 & [0.073, 0.104] & 0.091 & [0.069, 0.112] & 0.089 & [0.075, 0.105] & 0.063 & [0.047, 0.079] & 0.062 & [0.046, 0.078]\\
 \midrule
 CVL  &  &  &  &  &   & &  &  & &  \\
 \quad No (ref)& - & - &  -&-  & -  &-  &  &  & & \\
 \quad Yes  & 0.184 & [0.164, 0.203] & 0.163 & [0.137, 0.188] & 0.178 & [0.157, 0.198] & 0.198  & [0.178, 0.218] & 0.201& [0.181, 0.220]\\
 \midrule
 AL  &  &  &  &  &   &  &  &  & & \\
 \quad No (ref)& - & - &  -&-  & -  &-  &  &  & & \\
 \quad Yes  & 0.198 & [0.177, 0.219] & 0.177 & [0.149, 0.205] & 0.193 & [0.172, 0.213]  & 0.199 &[0.178, 0.220]  & 0.202& [0.181, 0.223]\\
 \midrule
 IPM  &  &  &  &  &   & &  &  & &  \\
 \quad No (ref)& - & - &  -&-  & -  &- &  &  & & \\
 \quad Yes  & 0.533 & [0.471, 0.596] & 0.476 & [0.401, 0.552] & 0.535 & [0.473, 0.599] & 0.544 & [0.481, 0.606] & 0.542& [0.479, 0.604]\\
 \midrule
 EMO  &  &  &  &  &   &  &  &  & & \\
 \quad No (ref)& - & - &  -&-  & -  &-  &  &  & & \\
 \quad Yes  & 0.912 & [0.822, 1.001] & 0.798 & [0.691, 0.906] & 0.908 & [0.817, 0.996] & 0.993 & [0.902, 1.083] & 0.994& [0.904, 1.085] \\
 \midrule
 IABP  &  &  &  &  &   &  &  &  & & \\
 \quad No (ref)& - & - &  -&-  & -  &-  &  &  & & \\
 \quad Yes  & 0.343 & [0.281, 0.404] & 0.297 & [0.216, 0.377] & 0.340 & [0.274, 0.398] & 0.324 & [0.261, 0.387] & 0.317& [0.254, 0.379]\\
 \midrule
  Age  &  &  &  &  &   & &  &  &   & \\
 \quad 18-39 (ref)& - & - &  -&-  & -  &- &  &  &   & \\
 \quad 40-80   & 0.133 & [0.107, 0.159] & 0.128 & [0.093, 0.162] & 0.130 & [0.102, 0.158] & 0.115 & [0.089, 0.142] &  0.116 & [0.089, 0.143]\\
 \quad $>80$ & 0.075 & [0.044, 0.106] & 0.063 & [0.022, 0.105] & 0.077 & [0.040, 0.110] & 0.057 & [0.026, 0.089] & 0.058  & [0.027, 0.090]\\\midrule
 Pre-LOS  &  &  &  &  &   & &  &  &   &\\
 \quad $\leq 1$ day (ref)& - & - &  -&-  & -  &- &  &  &   &\\
 \quad 2-7 days   & 0.038 & [0.017, 0.059] & 0.040 & [0.010, 0.069] & 0.038 & [0.016, 0.060]  &  0.040& [0.019, 0.062] &  0.040 & [0.019, 0.062]\\
 \quad $\geq 7$ days & 0.122 & [0.100, 0.145] & 0.140 & [0.109, 0.171] & 0.121 & [0.100, 0.143]  & 0.130 & [0.107, 0.153] & 0.129  & [0.106, 0.152]\\
\toprule
\end{tabular}
\end{center}
\end{table}
\end{landscape}

\begin{landscape}
\vspace*{\fill}
\begin{table}[!ht]
\begin{center}
\caption*{\textbf{TABLE 4 Continued} \textit{Results for ICU ventilation duration analysis:} Posterior means (Mean) with 95\% credible intervals (95\% Cred. Int.) from the VB algorithms and MCMC-based \textit{survregbayes}, respectively. Point estimates (Est.) with 95\% confidence interval (95\% Conf. Int.) from the h-likelihood \textit{survregHL} and \textit{survreg} methods.}
\begin{tabular}{llllllllllllllllll}
\toprule
\multicolumn{1}{l}{} & \multicolumn{6}{c}{With frailty} &  \multicolumn{4}{c}{Without frailty}\\
\cmidrule(rl){2-7} \cmidrule(rl){8-11} 
\multicolumn{1}{l}{} & \multicolumn{2}{c}{\textit{survregVBfrailty}} &  \multicolumn{2}{c}{\textit{survregHL}} & \multicolumn{2}{c}{\textit{survregbayes}}& \multicolumn{2}{c}{\textit{survreg}}& \multicolumn{2}{c}{\textit{survregVB}} \\
\cmidrule(rl){2-3} \cmidrule(rl){4-5} \cmidrule(rl){6-7}\cmidrule(rl){8-9}\cmidrule(rl){10-11}
 & Mean  & 95\% Cred. Int. & Est.  & 95\% Conf. Int. & Mean  & 95\% Cred. Int. & Est.  & 95\% Conf. Int. & Mean  & 95\% Cred. Int.\\\midrule
 MODS  &  &  &  &  &   & &  &  &   & \\
 \quad None (ref)& - & - &  -&-  & -  &-  &  &  &   &\\
 \quad Minimal  & 0.101 & [0.076, 0.127] & 0.112 & [0.076, 0.148] & 0.103 & [0.078, 0.130]  & 0.083 & [0.057, 0.109] & 0.082  & [0.056, 0.108]\\
 \quad Mild & 0.211 & [0.187, 0.235] & 0.228 & [0.191, 0.264] & 0.210 & [0.186, 0.233] & 0.182 & [0.158, 0.207] & 0.180  & [0.156, 0.205] \\
 \quad Moderate   & 0.265 & [0.235, 0.295] & 0.290 & [0.247, 0.332] & 0.262 & [0.231, 0.293]  & 0.248 & [0.217, 0.278] & 0.245 &[0.215, 0.276] &\\
 \quad Severe & 0.133 & [0.072, 0.195] & 0.198 & [0.122, 0.274] & 0.130 & [0.070, 0.189] & 0.137 & [0.075, 0.198] & 0.132  & [0.071, 0.193] \\
\toprule
\end{tabular}
\end{center}
\end{table}
\vspace*{\fill}
\end{landscape}
\end{document}